\documentclass[%
 preprint, linenumbers=false,
 amsmath,amssymb,
 aps, physrev,
longbibliography
]{revtex4-2}

\UseRawInputEncoding
\usepackage{graphicx}
\usepackage{dcolumn}
\usepackage{bm}
\usepackage{subcaption}
\usepackage{comment}
\usepackage{xcolor}

\begin{document}

\preprint{APS/123-QED}

\title{\textbf{Viscoplasticity can stabilise liquid collar motion on vertical cylinders} 
}%

\author{James D. Shemilt$^{1,2}$}\thanks{Contact author: jds74@cam.ac.uk}\author{Alice B. Thompson$^{1}$}\author{Alex Horsley$^{3}$}\author{Carl A. Whitfield$^{1,3}$}\author{Oliver E. Jensen$^{1}$}
\affiliation{%
$^{1}$Department of Mathematics, University of Manchester, Manchester, UK\\
$^{2}$Department of Applied Mathematics and Theoretical Physics, University of Cambridge, Cambridge, UK\\
$^{3}$Division of Immunology, Immunity to Infection and Respiratory Medicine, University of Manchester, Manchester, UK
}%


\begin{abstract}
Liquid films coating vertical cylinders can form annular liquid collars which translate downwards under gravity. We investigate the dynamics of a thin viscoplastic liquid film coating the interior or exterior of a vertical cylindrical tube, quantifying how the yield stress modifies both the Rayleigh-Plateau instability leading to collar formation and the translation of collars down the tube. We use thin-film theory to derive an evolution equation for the layer thickness, which we solve numerically to \textcolor{black}{examine} the nonlinear dynamics. 
\textcolor{black}{Instability and collar formation occur} when gravity is sufficiently strong to make the fluid yield \textcolor{black}{initially}. We use matched asymptotics to derive a model describing the quasi-steady translation of a slender liquid collar when the Bond number is small. The structure of the asymptotic solution for a viscoplastic collar shares some features with the Newtonian version, but there are several novel asymptotic regions that emerge at the two ends of the collar. The global force balance, which determines the collar's speed, is modified by a leading-order contribution from viscous drag in the collar when the liquid is viscoplastic. 
We use the asymptotic model to describe slow changes in collar volume when the film thicknesses ahead of, and behind, the collar are unequal. When the film thickness ahead of the collar is less than a critical value that we determine, viscoplastic collars adjust their volume and reach a steadily-translating state. This contrasts with the Newtonian problem, where the only state in which steady translation occurs is unstable to small changes in the film thickness. 
\end{abstract}

\maketitle


\section{Introduction}
\textcolor{black}{
A liquid film coating the interior of a cylindrical tube, or the exterior of a cylindrical tube or fibre, can be unstable to a surface-tension-driven Rayleigh-Plateau instability, which causes an initially thin uniform coating to form annular collars \cite{craster_dynamics_2009}. Droplets may form on the exterior of cylinders or liquid plugs within the interior of tubes, if the coating is thicker \cite{quere1999fluid,heil_mechanics_2008}. Motivation for studying the formation and motion of collars and plugs inside cylindrical tubes often arises from the application to modelling mucus mechanics in lung airways \cite{levy2014pulmonary,jensen_draining_2000,heil_mechanics_2008}. Airway mucus has a yield stress \cite{kavishvar2023yielding}, which is raised in diseases such as cystic fibrosis or chronic obstructive pulmonary disease (COPD) \cite{patarin_rheological_2020}. The yield stress can alter the dynamics of mucus transport in airways \cite{zamankhan_steady_2012,hu_microfluidic_2015,hu_effects_2020,shemilt2022surface,shemilt2023surfactant,erken_2022_elastoviscoplastic,fazla2024effects}, including potentially modifying the stresses exerted on the airway wall during flows of mucus; these stresses on the airway wall can lead to epithelial cell damage \cite{huh_acoustically_2007}. Understanding the impacts of yield stress on the gravity-driven flow of airway mucus provides motivation for this investigation into the effects of viscoplasticity on the formation and motion of liquid collars inside cylindrical tubes. Additional applications may be found in industrial and manufacturing processes in which complex fluids are coated on fibres or tubes, where formation of liquid collars or droplets may be disadvantageous \cite{quere1999fluid,craster_dynamics_2009}. }

\textcolor{black}{
Jensen \cite{jensen_draining_2000} developed a matched-asymptotics model for the gravity-driven translation of annular collars and plugs of Newtonian liquid on the interior of cylindrical tubes, when gravity is relatively weak compared to capillary effects. \textcolor{black}{The limit of weak gravity compared to surface tension is relevant when modelling the small conducting airways in the lungs \cite{jensen_draining_2000}. Jensen \cite{jensen_draining_2000} showed} that the asymptotic solution for a Newtonian collar (or plug) has a three-region structure, with short regions matching the main body of the collar (or plug) onto the uniform film ahead of, and behind, the collar (or plug). The speed of translation was determined via a force balance between the weight of the collar or plug and the viscous drag from the two short regions at the ends. A similar asymptotic structure has been identified in related problems where slender collars translate quickly under strong gravitational forces \cite{kalliadasis1994drop,chang1999mechanism,yu2013velocity}. Whilst the majority of studies have focused on the motion of Newtonian collars, droplets or plugs, Yu \& Hinch \cite{yu2014drops} also used matched asymptotics to study the fast translation of slender collars of power-law fluid, quantifying how the translation speed can depend on the power-law index. }

Formation of droplets or collars via a Rayleigh-Plateau instability has been observed in experiments in which fibres or rods are dip-coated from baths of Newtonian fluid (see the reviews by Craster \& Matar \cite{craster_dynamics_2009} or Qu\'er\'e \cite{quere1999fluid}). Recently, dip-coating of cylindrical rods in viscoplastic fluid has been investigated experimentally and theoretically \cite{smit2019stress,smit2021withdrawal}, but Rayleigh-Plateau instability was not reported. Lack of instability in those experiments may be due to the fluid yield stress or the rod diameter being too large for instability to occur in the liquid layer after it has coated the rod. Theoretical modelling, such as we present here, offers predictions that could subsequently be tested in new experiments in which either the interior of cylindrical tubes, or exterior of cylindrical rods or fibres, are coated with yield-stress fluids. 

The first investigation of the nonlinear evolution of a Newtonian liquid film coating the interior of a cylindrical tube was by Hammond \cite{hammond_nonlinear_1983}, whose model was later extended by Gauglitz \& Radke \cite{gauglitz_extended_1988} to study thicker films which may form plugs in the tube. Models similar to that of Gauglitz \& Radke \cite{gauglitz_extended_1988} have since been used to study the surface-tension-driven instability of Newtonian films in various scenarios, including where air flow also drives motion in the layer \cite{halpern_nonlinear_2003,camassa_2017_air}, in which surfactant is present at the interface \cite{halpern_surfactant_1993,ogrosky_linear_2021}, or where the liquid is viscoelastic \cite{halpern_effect_2010}. Numerical simulations using computational fluid dynamics (CFD) have also been used to study plug formation in axisymmetric Newtonian \cite{romano_liquid_2019,romano_2022_surfactant} and non-Newtonian layers \cite{romano_effect_2021,erken_2022_elastoviscoplastic,fazla2024effects}; these simulations are able to capture both the pre-coalescence and post-coalescence phases of plug formation. Erken et al. \cite{erken_2022_elastoviscoplastic} simulated plug formation using an elastoviscoplastic model for the liquid, with parameter values fitted to those from rheometry of airway mucus, and Fazla et al. \cite{fazla2024effects} later extended the model to incorporate kinematic hardening, showing that interactions between the effects of the different rheological properties in the model can modify the dynamics around plug formation. Camassa, Ogrosky \& Olander \cite{camassa_2014_gravity} conducted experiments in which a Newtonian liquid layer drained down the interior of a vertical tube, and compared results to a reduced-order long-wave model. Their model could capture formation of liquid collars or plugs, and it was found to predict well the critical film thicknesses for plug formation to occur. Dietze \& Ruyer-Quil \cite{dietze2020falling} also studied Newtonian films coating the interior of vertical cylindrical tubes in the case that the Bond number (a measure of the strength of gravitational forces compared to surface tension) is relatively large, and determined criteria for occlusion to occur via numerical simulations. Propagation and rupture of viscoplastic liquid plugs has also been studied both numerically and experimentally in the case that motion is driven by an imposed pressure gradient rather than by gravity \cite{zamankhan_steady_2012,hu_microfluidic_2015,hu_effects_2020}. In the present study, we focus on the formation and propagation of slender liquid collars at relatively low Bond numbers. We have previously conducted theoretical studies of the Rayleigh-Plateau instability in viscoplastic liquid layers coating cylindrical tubes, using thin-film and long-wave theories, in the case that there is no gravity \cite{shemilt2022surface,shemilt2023surfactant}. It was found that increasing the yield stress can have a significant stabilising effect, slowing growth of the instability and reducing the size of collars that form when the layer is thin. 

Viscoplastic thin-film theory, which will be used in this study, predicts that regions of plug-like flow develop in the liquid layer wherever the shear stress drops below the yield stress. These regions of plug-like flow typically develop adjacent to the free surface, and are termed `pseudo-plugs' \cite{walton_axial_1991} when the velocity within them varies along the direction of the flow, reflecting the fact that the fluid is marginally yielded within pseudo-plugs, with normal stresses and shear stresses combining to exceed the yield stress \cite{BALMFORTH199965}. Various free-surface flows of viscoplastic fluids, including flows driven by gravity, surface tension, or both, have been investigated previously using thin-film theory \cite{BALMFORTH199965,balmforth_visco-plastic_2000,balmforth_surface_2007,jalaal_stoeber_balmforth_2021,jalaal_long_2016,van_der_kolk_tieman_jalaal_2023,ball2024viscoplastic,shemilt2022surface,shemilt2023surfactant,balmforth2025implications}. \textcolor{black}{For example, Ball \& Balmforth \cite{ball2024viscoplastic} used thin-film theory to study the flow of a viscoplastic film coating the interior of a rotating horizontal cylinder, predicting how liquid pools at the base of the cylinder in the case that flow is primarily driven by gravity and surface tension plays a relatively minor role.} Predictions from thin-film theory have been directly compared to computational fluid dynamics (CFD) simulations or experiments in a number of studies of surface-tension-driven flows \cite{jalaal_long_2016,jalaal_stoeber_balmforth_2021,van_der_kolk_tieman_jalaal_2023}. In a study of steady bubble propagation through a tube filled with viscoplastic fluid, Jalaal \& Balmforth \cite{jalaal_long_2016} utilised both a thin-film model and CFD simulations, finding generally good agreement when the liquid film between the bubble and the tube wall was thin. Jalaal et al. \cite{jalaal_stoeber_balmforth_2021} also found good agreement between thin-film predictions and CFD results for the spreading radius of viscoplastic droplets, except when the droplet shape deviated significantly from the shallow profile assumed in their thin-film theory. In their numerical simulations, they also identified regions adjacent to the free surface of the droplets resembling weakly-yielded pseudo-plugs when the droplets were sufficiently shallow, validating the flow structure predicted by the thin-film theory. \textcolor{black}{At the leading edge of a spreading droplet \cite{jalaal_stoeber_balmforth_2021} or at the edge of the end-cap of a translating bubble in a tube \cite{jalaal_long_2016}, there are capillary undulations at the liquid's free surface, which coincide with rapid changes in the sign of the shear stress and direction of flow. It is well-established that viscoplastic thin-film theory breaks down in asymptotically small regions around points where the shear stress changes sign \cite{balmforth_visco-plastic_2000}, but comparisons with CFD simulations in studies such as \cite{jalaal_stoeber_balmforth_2021} and \cite{jalaal_long_2016} suggests the theory can still provide accurate predictions of shallow surface-tension-driven flows. }

In this study, we employ thin-film theory to investigate the Rayleigh-Plateau instability of a viscoplastic liquid film coating a vertical cylinder and the motion of slender liquid collars on the cylinder. We aim to quantify the effects of viscoplasticity on the dynamics of liquid collar formation and propagation. By developing a small-Bond-number model for quasi-steady collar motion, we make theoretical predictions for how the liquid's yield stress modifies the speed, size and stability of a translating collar. 
The paper is organised as follows. In \S\ref{sec:model_form}, we derive a thin-film evolution equation for the flow in the liquid layer, and briefly discuss methods for solving it numerically. We present results for the evolution of \textcolor{black}{initially almost flat} films in \S\ref{sec:results_IVP}, with \textcolor{black}{discussion of how instability can be triggered by sinusoidal free-surface perturbations} in \S\ref{sec:results_LSA} and an illustrative example of the nonlinear evolution of a layer presented in \S\ref{sec:results_nonlinear}. In \S\ref{sec:asymp_model}, we derive a model for quasi-steadily translating collars using matched asymptotics, detailing the solution in each of the asymptotic regions sequentially. In doing so, we extend and modify the approach used in previous studies of Newtonian fluid collars \cite{jensen_draining_2000}. We present results from the asymptotic model in \S\ref{sec:results_smallB}, focusing on steadily-translating collars in \S\ref{sec:results_steady}, where we also present comparisons between asymptotic and numerical results, and focusing on the asymptotic model's predictions for the unsteady dynamics of translating collars presented in \S\ref{sec:results_unsteady}.

\section{Model formulation}\label{sec:model_form}


\subsection{Governing equations}

\begin{figure*}[t!]
    \begin{subfigure}{0.47\textwidth}
        \centering
        \includegraphics[width = \textwidth]{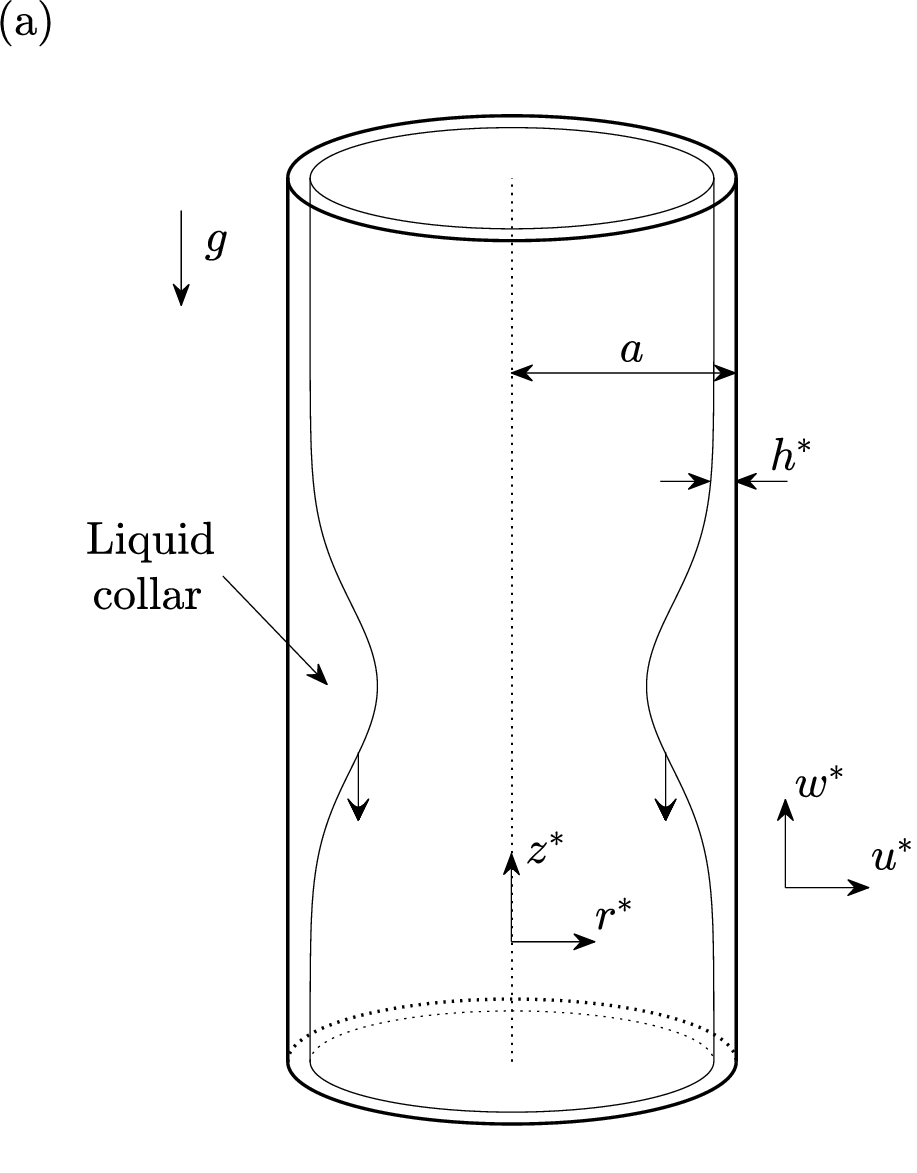}
    \end{subfigure}%
    \hfill
    \begin{subfigure}{0.48\textwidth}
        \centering
        \includegraphics[width = \textwidth]{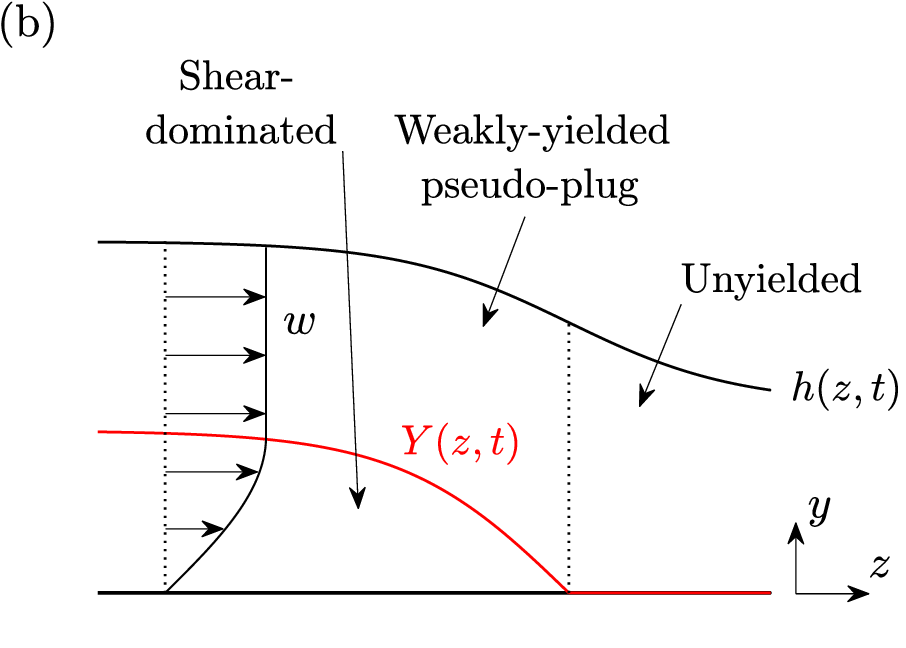}
        \hfill
        \vspace{40pt}
    \end{subfigure}
    \caption{(a) Sketch of the model geometry. (b) Typical flow structure within the thin liquid film. In dimensionless coordinates, $y=h$ is the free surface, $y=Y$ is the boundary between regions of shear-dominated flow and weakly-yielded pseudo-plugs, and $w$ is the axial velocity. }
    \label{fig:geomsketch}
\end{figure*}

We consider a vertical cylindrical tube of radius $a$, lined on the inside by a layer of viscoplastic liquid (figure \ref{fig:geomsketch}a). The rest of the tube is filled with an inviscid gas, in which the pressure is assumed to be uniform. We assume that flow in the liquid layer is axisymmetric, so the system can be described by coordinates $(r^*,z^*)$, where the tube wall lies at $r^*=a$, and gravity acts in the negative $z^*$-direction. We take the gas-liquid interface to lie at $r^* = a - h^*(z^*,t^*)$. We assume that there is no inertia in the liquid layer and that it is incompressible, so the flow is governed by
\begin{eqnarray}
    0 &=& \partial_{z^*}w^* + \frac{1}{r^*}\partial_{r^*}(r^*u^*),\label{stokesmasscons}\\
    0 &=& -\partial_{r^*}p^* + \frac{1}{r^*}\partial_{r^*}(r^*\tau_{rr}^*)+\partial_{z^*}\tau_{rz}^*-\frac{\tau^*_{\theta\theta}}{r^*},\label{stokesvertmom}\\
    0 &=& -\partial_{z^*}p^* + \frac{1}{r^*}\partial_{r^*}(r^*\tau_{rz}^*)+\partial_{z^*}\tau_{zz}^* - \rho g,
    \label{stokeshorizmom}
\end{eqnarray}
where $(u^*,w^*)$ is the fluid velocity, $p^*$ is the pressure measured relative to the gas pressure, $\boldsymbol{\tau}^*$ is the deviatoric stress tensor, and $\rho g$ is constant. We assume that the liquid is a Bingham fluid, with constitutive relation,
\begin{equation}
\left. \begin{array}{ll}  
\displaystyle \tau_{ij}^* = \left(\eta + \frac{\tau_Y}{\dot\gamma^*}\right)\dot\gamma_{ij}^*
  &\quad \mbox{if\ } \tau^* > \tau_Y,\\[8pt]
\displaystyle  \dot\gamma_{ij}^* = 0
  &\quad \mbox{if\ } \tau^*\leq\tau_Y,
 \end{array}\right\}
  \label{stokesconstit}
\end{equation}
where $\eta$ is a viscosity, $\tau_Y$ is the yield stress, $\boldsymbol{\dot\gamma}^*$ is the strain-rate tensor, and $\dot\gamma^*$ and $\tau^*$ are the second invariants of strain-rate and deviatoric stress, respectively. 

We assume no slip at the cylinder wall, so
\begin{equation}
    u^* = w^* = 0 \quad\mbox{at}\quad r^* = a.\label{noslipBC}
\end{equation}
At the free surface, the kinematic and stress conditions are
\begin{eqnarray}
    \partial_{t^*}h^* + w^* \partial_{z^*}h^* = -u^* \quad &\mbox{on\ }&\quad r^*=a - h^*,\label{stokeskinBC}\\
    -p^*n_i + \tau^*_{ij}n_j = \sigma\kappa^* n_i \quad &\mbox{on\ }&\quad r^*=a - h^*,
    \label{stressBC}
\end{eqnarray}
where $\boldsymbol{n}$ is the unit normal to the free surface, $\sigma$ is the constant surface tension, and the free-surface curvature is
\begin{equation}
    \kappa^* = \frac{1}{\sqrt{1+(\partial_{z^*}h^*)^2}}\left[\frac{1}{a-h^*}+\frac{\partial_{z^*z^*}h^*}{1+(\partial_{z^*}h^*)^2}\right].
\end{equation}

We consider two different scenarios, distinguished by the boundary conditions and initial conditions that we impose in each case. In both cases, we suppose that the layer has a characteristic thickness, $h^*_0$. The first scenario, which we discuss in \S\ref{sec:results_IVP} as well as in \S\ref{sec:results_steady}, is an initially flat layer, with
\begin{equation}
    h^*(z^*,t^*=0) = h_0^*,\label{hIC}
\end{equation}
in a periodic domain, $0\leq z^* \leq\mathcal{L}^*$. \textcolor{black}{In the second scenario, detailed in \S\ref{sec:asymp_model}, we assume that there is an isolated collar of liquid in a long tube, with the film thickness taking a constant value, $h^*=h^{*-}_\infty$, ahead of the collar and another value, $h^*=h^{*+}_\infty$, far behind the collar. We derive a model for the slow time evolution of the collar in the limit $B\rightarrow0$, where $B$ is a Bond number that we define below. The far-field film thicknesses, $h^{*\pm}_\infty$, are assumed to be much smaller than the characteristic thickness, $h^*_0$, of the collar itself. When solving for the time evolution of the isolated collar, we take $h_0^* = V^*_0/(16\pi a^2)$, where $V_0^*$ is the initial volume of the collar, excluding the volume of the very thin films ahead of and behind the collar. The exact choice of $h_0^*$ is not of unique importance, except that $h_0^*\ll a$; rescaling $h_0^*$ by a constant $O(1)$ factor would not affect the results in \S\ref{sec:asymp_model}.} 

\subsection{Non-dimensionalisation and thin-film scaling}

We consider the system \eqref{stokesmasscons}-\eqref{hIC} in a thin-film limit, in which $\epsilon \equiv h^*_0/a \ll 1$. We non-dimensionalise and introduce the thin-film scaling by defining
\begin{equation}
\left. \begin{array}{l}
\displaystyle
( y, z) = \left(\frac{a - r^*}{\epsilon a},\frac{z^*}{a}\right),\quad (u,w) = \frac{\eta}{\sigma}\left(-\frac{u^*}{\epsilon^4},\frac{w^*}{\epsilon^3}\right), \quad h = \frac{h^*}{\epsilon a}, \quad\boldsymbol{  \tau} = \frac{a}{\epsilon^2\sigma}\boldsymbol{\tau}^*, \\[16pt]
\displaystyle

        \boldsymbol{{\dot\gamma}}=\frac{\eta a}{\epsilon^2\sigma}\boldsymbol{\dot\gamma}^*,  \quad {t}= \frac{\epsilon^3\sigma}{a\eta}t^*,\quad  -1 + \epsilon p= \frac{a}{\sigma}p^*, \quad   1 + \epsilon\kappa = a\kappa^*, \quad \mathcal{L} = \frac{\mathcal{L}^*}{a},
\end{array} \right\}
\label{nondim}
\end{equation}
before truncating at leading order in $\epsilon$. After truncating, \eqref{stokesmasscons}-\eqref{stokeshorizmom} become
\begin{eqnarray}
    0 &=&\partial_yu + \partial_zw ,\label{TFmasscons}\\
    0 &=& \partial_yp,\label{TFvertmom}\\
    0 &=& -\partial_zp + \partial_y\tau_{yz} - B,\label{TFhorizmom}
\end{eqnarray}
where 
\begin{equation}
    B = \frac{\rho ga^2}{\epsilon\sigma}
    \label{Bdefn}
\end{equation}
is the Bond number. The stress boundary conditions at the free surface \eqref{stressBC} become
\begin{eqnarray}
    p = -\kappa \quad\mbox{and}\quad \tau_{yz} = 0 \quad \mbox{at}\quad y = h,\label{TFstressBCs}
\end{eqnarray}
where
\begin{equation}
    \kappa = h + \partial_{zz}h.\label{TFkappa}
\end{equation}
Integrating \eqref{TFhorizmom}, and using the boundary condition \eqref{TFstressBCs}, gives the shear stress
\begin{equation}
    \tau_{yz} = (\partial_zp+B)(y-h),
    \label{TFtauyz}
\end{equation}
and integrating \eqref{TFvertmom}, with \eqref{TFstressBCs}, gives $p = -\kappa$. We define the wall shear stress,
\begin{equation}
    \tau_\mathrm{w} = \tau_{yz}|_{y=0} = -h(B - \partial_{z}h - \partial_{zzz}h).\label{wallstressdefn}
\end{equation}
The constitutive law \eqref{stokesconstit}, after non-dimensionalising, becomes
\begin{equation}
\left. \begin{array}{ll}  
\displaystyle \tau_{ij} = \left(1 + \frac{J}{\dot\gamma}\right)\dot\gamma_{ij}
  &\quad \mbox{if\ } \tau > J,\\[8pt]
\displaystyle  \dot\gamma_{ij} = 0
  &\quad \mbox{if\ } \tau\leq J,
 \end{array}\right\}
  \label{TFconstit}
\end{equation}
where
\begin{equation}
    J = \frac{\tau_Ya}{\epsilon^2\sigma}.
\end{equation}
is the capillary Bingham number.

The structure of the flow resembles other thin-film viscoplastic flows \cite{BALMFORTH199965} and so the derivation of the axial velocity, $w$, proceeds essentially as in \cite{shemilt2022surface}. We define a surface, $y=Y(z,t)$, such that $|\tau_{yz}|=J$ on $y=Y$ if $\tau_\mathrm{w}>J$, and $Y=0$ if $\tau_\mathrm{w}\leq J$. 
In $0\leq y \leq Y$, the flow is shear-dominated, with the normal stresses asymptotically smaller than the shear stress, so \eqref{TFtauyz} can be combined with \eqref{TFconstit} and integrated to give the leading-order expression for $w$. In $Y< y \leq h$, there is a `pseudo-plug', a region of plug-like flow where $w = w_\mathrm{p}(z,t)$ to leading order. 
Wherever $Y=0$, the axial velocity is zero, $w=0$, and the fluid can be considered to be unyielded. Figure \ref{fig:geomsketch}(b) illustrates the typical structure of the thin-film flow. Combining \eqref{TFtauyz} and \eqref{TFconstit}, integrating in $y$ and using the no-slip boundary condition \eqref{noslipBC}, we find that the leading-order axial velocity is
\begin{equation}
    w = \left\{
    \begin{array}{ll}
      \frac{1}{2}(B-h_z-h_{zzz}) y(y-2Y), & \quad \mbox{for}\quad 0\leq y < Y \\[2pt]
      w_\mathrm{p} \equiv -\frac{1}{2}(B-h_z-h_{zzz})Y^2, &\quad \mbox{for}\quad Y \leq y \leq h,
    \end{array} \right.\label{TFvelocity}
\end{equation}
where, from now on, subscripts $z$ or $t$ denote partial derivatives with respect to those variables. The kinematic condition \eqref{stokeskinBC}, after non-dimensionalising and combining with \eqref{noslipBC} and \eqref{TFmasscons}, gives the evolution equation, 
\begin{equation}
    h_t + \left[w_{\mathrm{p}}\left(h-\frac{Y}{3}\right) \right]_z = 0 \quad \mbox{where} \quad Y = \mathrm{max}\left\{0,\,h\left(1-\frac{J}{|\tau_\mathrm{w}|}\right)\right\},\label{TFevoleqn}
\end{equation}
where the mass flux in the square bracket in \eqref{TFevoleqn} was determined by integrating \eqref{TFvelocity} across $0\leq y \leq h$, and the expression for $Y$ follows from \eqref{TFtauyz} and the definition \eqref{wallstressdefn}. The initial condition \eqref{hIC} becomes
\begin{equation}
    h(z,t=0) = 1.\label{TFIC}
\end{equation}
Whilst we have derived \eqref{wallstressdefn}, \eqref{TFvelocity} and \eqref{TFevoleqn} assuming the liquid layer is coating the interior of the cylinder, the same process can be used to derive the evolution equation for a thin film coating the exterior of a cylinder. 

\subsection{Numerical methods}

To solve \eqref{TFevoleqn} numerically, we employed the regularisation introduced by Jalaal \cite{jalaal_thesis_2016}, which has also been used in various other thin-film viscoplastic flows \cite{jalaal_long_2016,jalaal_stoeber_balmforth_2021,shemilt2022surface,shemilt2023surfactant}. We introduce the regularisation parameter, $Y_\mathrm{min}$, and replace $Y$ in \eqref{TFevoleqn} by $\bar{Y}=\max(Y_\mathrm{min},Y)$. This regularisation introduces weak flow where the thin-film theory predicts no flow, but does not affect the evolution equation where $Y>Y_\mathrm{min}$. We use $Y_\mathrm{min}=10^{-6}$, after having checked that this value is small enough that results are independent of the exact value. We solve the regularised version of \eqref{TFevoleqn} using second-order centred finite differences to approximate the spatial derivatives, and an ODE solver in \textsc{Matlab} to solve the resulting set of equations through time. The number of finite-difference grid points used varied between $400$ and $1000$. Where fewer grid points were used, it was checked that results were not affected in any significant way by the exact number of grid points used. In \S\ref{sec:results_LSA}, we \textcolor{black}{discuss the stability of an almost flat layer} with no regularisation of the evolution equation, and in the small-$B$ model derived in \S\ref{sec:asymp_model}, there is also no regularisation of the equations.

\section{Instability and time evolution of a {viscoplastic film} }\label{sec:results_IVP}


\subsection{\textcolor{black}{Instability of an almost flat layer}}\label{sec:results_LSA}

\color{black}
In this section, we consider the stability of an almost flat layer, a uniform film that has been subject to some perturbation in the free-surface height, $h(z,t)$. For an exactly uniform film \eqref{TFIC}, from the definition of $Y$ in \eqref{TFevoleqn} the fluid is flowing if and only if $B>J$, or, equivalently, $\mathcal{J}<1$, where we define 
\begin{equation}
    \mathcal{J}= \frac{J}{B}.\label{Jscaleddefn}
\end{equation}
For the uniform film, the fluid is fully yielded in $0\leq y \leq Y = 1-\mathcal{J}$. 

When $\mathcal{J}=0$, a uniform film is known to be linearly unstable to long-wave perturbations \cite{hammond_nonlinear_1983}. When $0<\mathcal{J}<1$, if the layer height is exactly uniform, we expect there to be an unyielded plug adjacent to the free surface. For unstable growth to occur, the perturbation applied to the layer must be strong enough to yield the fluid at interface. A small perturbation to the stress would not, in general, be sufficient to break the plug and induce instability, but a perturbation to the velocity field could raise the stress above the yield stress and yield the layer, as discussed by Balmforth \& Liu \cite{balmforth2004roll} for the related problem of inclined plane flow. Whether there is a plug at the interface after a perturbation is applied to $h$ may depend on the amplitude of that perturbation. 

The thin-film theory employed in \S\ref{sec:model_form} assumes that pseudo-plugs exist wherever the magnitude of the shear stress drops below the yield stress, and that these pseudo-plugs are able to deform. Liu et al. \cite{liu_viscoplastic_2019} demonstrated that, for gravity currents flowing down inclines, where the free-surface height deviates from uniform by an amount of the order of $\epsilon$, the thin-film parameter, then a solution can exist with a genuinely rigid plug at the free surface. An alternative thin-film theory with a rigid plug at the interface is then required, which is combined with standard thin-film theory in regions where the free-surface height deviates from uniform by more than $O(\epsilon)$ \cite{liu_viscoplastic_2019}. Analogously, in the problem we consider here, we expect that the perturbation applied to the free-surface height initially must be large enough that the fluid adjacent to the free surface is a marginally-yielded pseudo-plug, for the thin-film evolution equation \eqref{TFevoleqn} to hold. In the following, we assume that the layer is perturbed in such a way that the fluid at the interface is yielded and \eqref{TFevoleqn} holds. As in \cite{liu_viscoplastic_2019}, an alternative theory may be required to investigate the behaviour of a flat layer subjected to perturbations with amplitude of $O(\epsilon)$ or smaller, taking into account the possibility of a rigid plug at the interface; we do not pursue this here, instead focusing on the finite-amplitude instability and dynamics described by \eqref{TFevoleqn}. Viscoplastic thin-film theory is also known to break down in asymptotically small regions around points where the direction of flow changes \cite{balmforth_visco-plastic_2000}, so it is possible that small rigid plugs may also exist in these regions that are not predicted by the theory; we do not examine these asymptotically small regions in detail, but assume that the thin-film theory still provides a good approximation to the global dyanmics.

To examine the instability of an almost flat layer, we assume that $\mathcal{J}<1$ and consider a sinusoidally perturbed free surface, 
\begin{equation}
	h(z,t) = 1 + Ae^{ikz + st},
\label{LSA:hpert}
\end{equation}
where $\epsilon \ll |A| \ll 1$. Substituting \eqref{LSA:hpert} into \eqref{TFevoleqn} gives the perturbed surface separating the fully-yielded region and pseudo-plug,
\begin{equation}
Y \sim 1 - \mathcal{J} + \left[1 - \frac{\mathcal{J}ik(1-k^2)}{B}\right]Ae^{ikz+st} + O(|A|^2),
\end{equation}
and the pseudo-plug velocity,
\begin{equation}
w_{\mathrm{p}} \sim -\frac{B}{2}(1-\mathcal{J})^2 - \frac{1}{2}(1-\mathcal{J})\left[2B - {i}k(1+\mathcal{J})(1-k^2)\right]Ae^{ikz+st} + O(|A|^2).
\end{equation}
After linearising, \eqref{TFevoleqn} gives
\color{black}
\begin{equation}
    s = \frac{1}{3}k^2(1-k^2)(1-\mathcal{J}^3) + ikB(1-\mathcal{J}) \quad\mbox{for}\quad 0\leq\mathcal{J}<1.\label{LSAdispersion}
\end{equation}
The Newtonian growth rate \cite{hammond_nonlinear_1983} is recovered by setting $\mathcal{J}=0$ in \eqref{LSAdispersion}. The growth rate, $\mathrm{Re}(s)$ in \eqref{LSAdispersion}, decreases monotonically as $\mathcal{J}$ is increased, and approaches zero as $\mathcal{J}\rightarrow1$. The deviation from the Newtonian growth rate in \eqref{LSAdispersion} is proportional to $\mathcal{J}^3$, so is small unless $\mathcal{J}\approx1$; viscoplasticity has a minor impact on the growth rate unless the yield stress is large enough that the fluid is almost entirely unyielded in the initial state. As in the Newtonian problem \cite{hammond_nonlinear_1983}, the system is unstable to disturbances with wavenumber $0<k<1$, and the most unstable wavelength corresponds to $k=k_\mathrm{m}\equiv\sqrt{2}/2$\textcolor{black}{, although, as discussed above, this argument holds in the viscoplastic case only when the free-surface perturbations are large enough to yield the fluid at the interface.}

\subsection{Nonlinear evolution}\label{sec:results_nonlinear}

\begin{figure*}[t!]
    \centering
    \begin{subfigure}[t]{\textwidth}
        \centering
        \includegraphics[width = \textwidth]{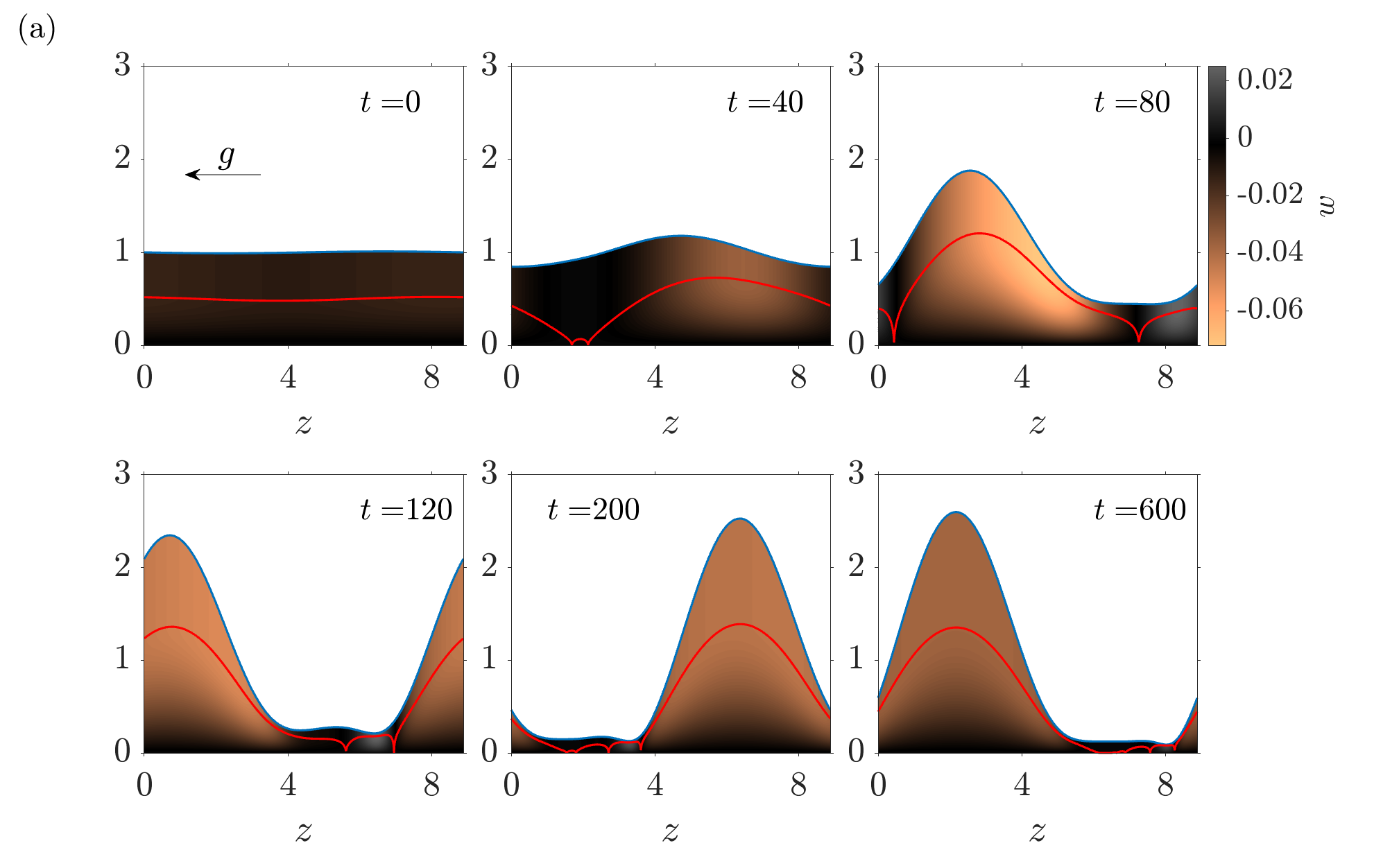}
    \end{subfigure}%
    \\
    \vspace{6pt}
    \begin{subfigure}{0.48\textwidth}
        \centering
        \includegraphics[width = \textwidth]{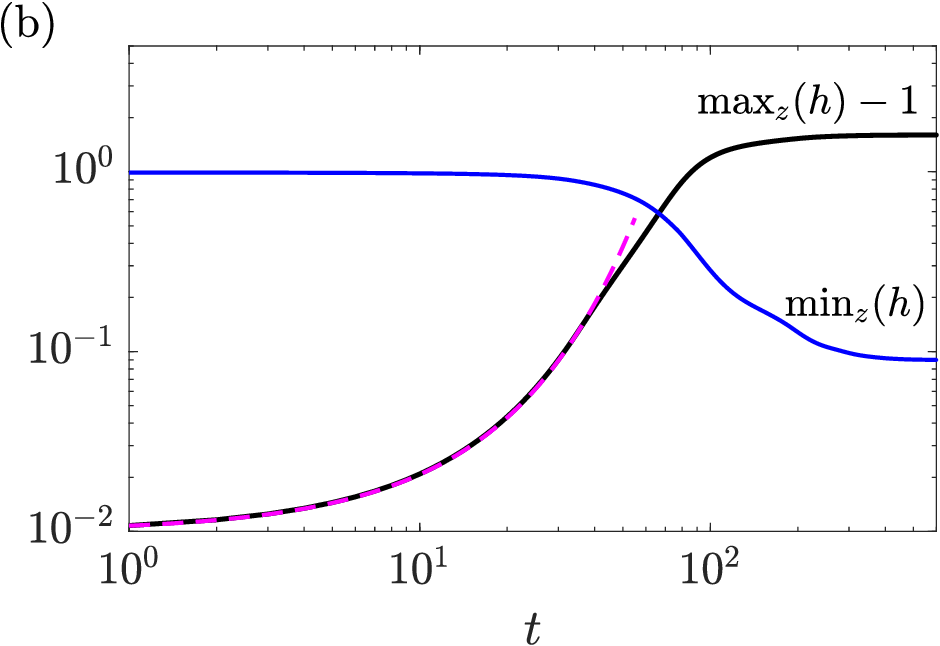}
    \end{subfigure}%
    \hfill
    \begin{subfigure}{0.49\textwidth}
        \centering
        \includegraphics[width = \textwidth]{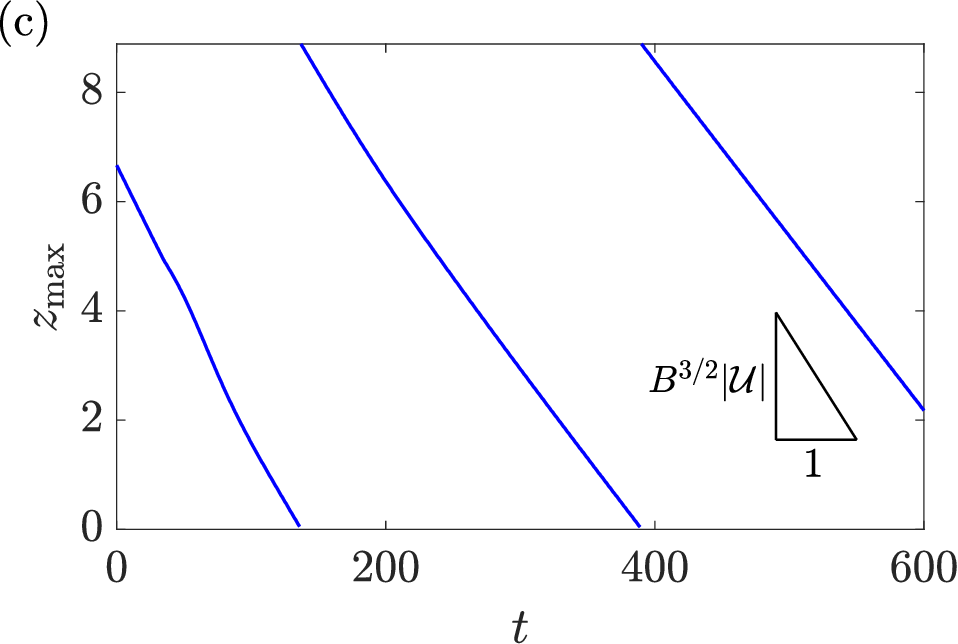}
    \end{subfigure}
    \caption{Numerical solution of \eqref{TFevoleqn} with $B=0.1$, $J=0.05$, in a periodic domain of length $\mathcal{L}=2\pi/k_\mathrm{m}=2\sqrt{2}\pi$. (a) Snapshots of the evolving layer. Lines show $h(z,t)$ (blue) and $Y(z,t)$ (red); colour map illustrates the axial velocity. Arrow in upper-left panel indicates the direction of gravity. (b) \textcolor{black}{Time evolution of $\max_z(h)-1$ (black) with the predicted \textcolor{black}{early-time} exponential growth rate from \eqref{LSAdispersion} (dashed magenta), and $\min_z(h)$ (blue).} (c) Location, $z_\mathrm{max}$, of the maximum of $h(z,t)$. The leading-order asymptotic prediction \eqref{res:steady_U} for the steady collar speed, $B^{3/2}|\mathcal{U}|$, is also illustrated in (c), where we have used the approximation $V=\mathcal{L}$ to calculate $\mathcal{U}$ in \eqref{res:steady_U}. }
    \label{fig:nonlinear_evolution}
\end{figure*}

To analyse the nonlinear evolution \textcolor{black}{of an initially almost flat layer}, we solve \eqref{TFevoleqn} numerically in a periodic domain, $0\leq z \leq \mathcal{L}$, with initial condition \eqref{TFIC} replaced by
\begin{equation}
    h(z,0) = 1 - 0.01\sin\left(\frac{2\pi z}{\mathcal{L}}\right).\label{numIC}
\end{equation}
These numerical simulations in fixed periodic domains can capture both the initial growth of the instability from a near-flat layer, leading to collar formation, and the subsequent translation of collars through the domain. In all numerical simulations presented, we set $\mathcal{L} = 2\pi/k_\mathrm{m} = 2\pi\sqrt{2}$, which means that the domain can accommodate one unstable mode, the most unstable mode from \eqref{LSAdispersion}. We have also run simulations in domains of different lengths, finding qualitatively similar results in general, except that when the domain is made significantly long more than one collar may develop. We will see below that collars have length of approximately $2\pi$ when $B$ is relatively small, so we expect that up to $n$ collars may develop in a domain of length $2n\pi<\mathcal{L}<2(n+1)\pi$. We focus on the emergence and evolution of a single collar, and so choose $\mathcal{L} = 2\pi/k_\mathrm{m}$, which satisfies $2\pi<\mathcal{L}<4\pi$. \textcolor{black}{It may be possible that in longer domains, multiple collars may interact and there may be secondary instabilities; we do not pursue this here, instead focusing on the dynamics of individual collars. }

Figure \ref{fig:nonlinear_evolution} shows output from a typical numerical simulation where instability and collar formation occur. In the initial state (figure \ref{fig:nonlinear_evolution}a, $t=0$), $Y\approx1-\mathcal{J}$ so there is flow in the layer but the velocity is relatively low throughout. After a short time ($t=40$), the free surface has become non-uniform and then ($t=80$) an emerging fluid collar is evident, which is translating in the negative $z$-direction. The growth rate of the peak height of the layer is well-approximated by the exponential growth rate predicted in \eqref{LSAdispersion} (figure \ref{fig:nonlinear_evolution}b). As the collar is growing, the axial velocity, $w$, is largest in the rear half of the collar (figure \ref{fig:nonlinear_evolution}a, $t=80$), where capillary forces and gravity are both driving fluid in the negative $z$-direction, whilst at the front of the collar, the velocity is weaker since capillary forces are acting in the opposite direction to gravity. At $t=80$, there are two points in the domain at which $Y\approx0$; these are points at which the direction of flow changes, indicating that at the front of the emerging collar there is a weak flow in the positive $z$-direction. Subsequently, the collar gradually grows, and a thin film develops in the rest of the domain. 

By $t=600$, the collar has essentially stopped growing (figure \ref{fig:nonlinear_evolution}b), and is translating at an almost constant speed (figure \ref{fig:nonlinear_evolution}c). From figure \ref{fig:nonlinear_evolution}(a), at $t=600$, there is a short region within the thin film, at around $z\approx6$, where $Y=0$ and $w=0$. For larger values of $B$, the film between the front and back of the collar would be thicker and more likely to be yielded throughout, but for the relatively small values of $B$ that we focus on, it is typical that there is an unyielded region, as in figure \ref{fig:nonlinear_evolution}(a) at $t=600$. \textcolor{black}{The minimum layer thickness approaches a non-zero value at late times (figure \ref{fig:nonlinear_evolution}b), confirming that a liquid film is continually deposited behind the collar.} At the front edge of the collar, $h$ undulates and there are several bumps in $Y$ (figure \ref{fig:nonlinear_evolution}a, $t=200, 600$), indicating that the direction of flow is changing repeatedly, although the velocity is low throughout this region. 
Predicting the late-time shape of $h$ and $Y$ is the focus of the modelling in \S\ref{sec:asymp_model}, and in \S\ref{sec:results_smallB} we compare predictions from the asymptotic model to late-time output from numerical simulations. 

\section{Small-$B$ model for quasi-steadily translating collars}\label{sec:asymp_model}

In this section, we use matched asymptotics to derive a model for the translation of a slender collar of liquid down the tube wall, considering \eqref{TFevoleqn} in the limit $B\ll1$. \textcolor{black}{Focusing on the limit $B\ll1$ enables us to develop an asymptotic model that will provide a detailed picture of the structure of a translating collar. We will subsequently compare results to numerical solutions of the evolution equation \eqref{TFevoleqn}, which is also valid when $B$ is not small. The analysis in this section} has some key differences from that of the Newtonian analogue \cite{jensen_draining_2000}. We consider an isolated collar of liquid translating with dimensionless speed $U(t)$, with a film of uniform thickness, $h_\infty^-$, ahead of the collar, and a film of uniform thickness, $h_\infty^+$, deposited behind the collar. We no longer assume that the domain is periodic, as we did in \S\ref{sec:results_IVP}, in order to describe general quasi-steady motion. We assume that $h=O(1)$ in the main body of the collar, but that the films ahead and behind are much thinner, so that $h_\infty^\pm\ll1$. We will consider the regime in which $B$, as defined in \eqref{Bdefn}, is of the same order of magnitude as the film thickness ahead of the collar, so $B= O(h_\infty^-)$, and in which the deposited film behind the collar is also of the same order of magnitude, $h_\infty^+ = O(B)$, although we do not, in general, enforce that $h_\infty^+=h_\infty^-$. We write $h_\infty^-=BH_\infty^-$ and $h_\infty^+=BH_\infty^+$, where $H_\infty^\pm=O(1)$, and treat $B$ as the small parameter in the problem. We could equally treat $h_\infty^-$ as the small parameter and treat $B$ as a function of $h_\infty^-$ (e.g., \cite{jensen_draining_2000}).

Motivated by numerical solutions to the evolution equation \eqref{TFevoleqn} such as figure \ref{fig:nonlinear_evolution}(c), we let $U = B^{3/2}\mathcal{U}$, assuming $\mathcal{U}=O(1)$. Given the thinness of the film ahead of the collar, we expect that the time scale for $O(1)$ changes to occur in the collar height to scale like $1/(h_\infty^-U)$, so we define $\mathcal{T} = B^{5/2}t$, where we assume $\mathcal{T}=O(1)$, which allows us to consider quasi-steady motion of the collar. We assume that the capillary Bingham number is of the same order of magnitude as the Bond number, setting $\mathcal{J}=O(1)$, with $\mathcal{J}$ defined as in \eqref{Jscaleddefn}; this ensures that the fluid can yield, and that the yield stress plays a role at leading order. There are also alternative regimes in which $J\ll B$, where viscoplasticity may still influence the dynamics, albeit to a lesser extent. We discuss one such regime, in which $J=O(B^{3/2})$, in appendix \ref{sec:appJB32}. 

We consider the system in a frame of reference moving with the velocity of the collar, so that the tube is translating with speed $-U\boldsymbol{\hat{z}}$ in the moving frame. Defining $\zeta$ to be the axial coordinate in the moving frame, \eqref{TFevoleqn} is modified to give
\begin{equation}
    B^{5/2}h_\mathcal{T} + q_\zeta = 0
    \quad \mbox{where} \quad
    q = {w}_\mathrm{p}\left(h-\frac{Y}{3}\right) - B^{3/2}\mathcal{U}h,
    \label{travelling_evoleqn}
\end{equation}
with the pseudo-plug velocity \eqref{TFvelocity},
\begin{equation}
      {w}_\mathrm{p} = -\frac{1}{2}(B-h_\zeta-h_{\zeta\zeta\zeta})Y^2,
      \label{asymp_wp}
\end{equation}
the boundary between fully-yielded fluid and pseudo-plugs \eqref{TFevoleqn},
\begin{equation}
    Y = \mathrm{max}\left\{0,h\left(1-\frac{B\mathcal{J}}{|\tau_\mathrm{w}|}\right)\right\},\label{YsmallBdefn}
\end{equation}
and the wall shear stress \eqref{wallstressdefn},
\begin{equation}
      \tau_\mathrm{w} = -h(B-h_\zeta-h_{\zeta\zeta\zeta}),
      \label{asymp_tauw}
\end{equation}
all essentially unchanged. 

\begin{figure*}[t!]
    \centering
    \begin{subfigure}[t]{\textwidth}
        \centering
        \includegraphics[width = 0.94\textwidth]{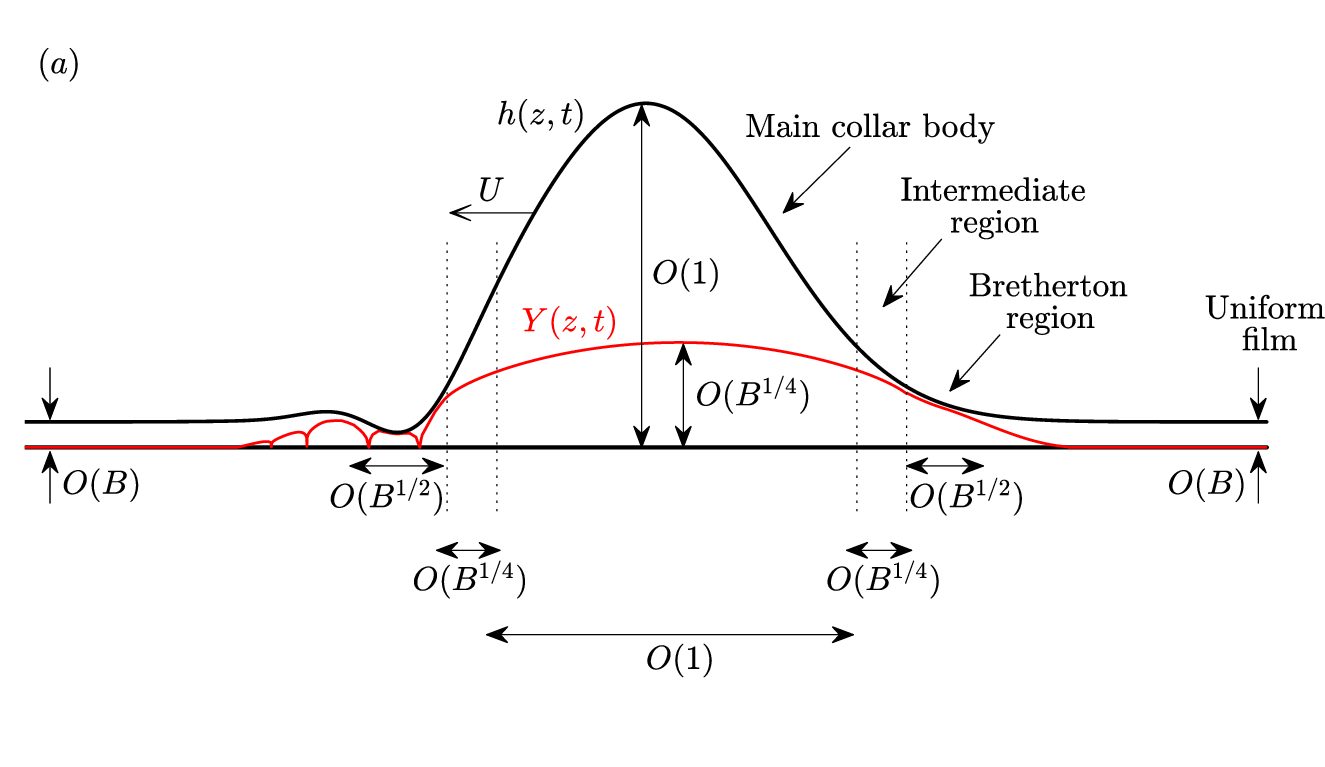}
    \end{subfigure}%
    \\
    \begin{subfigure}{0.48\textwidth}
        \centering
        \includegraphics[width = \textwidth]{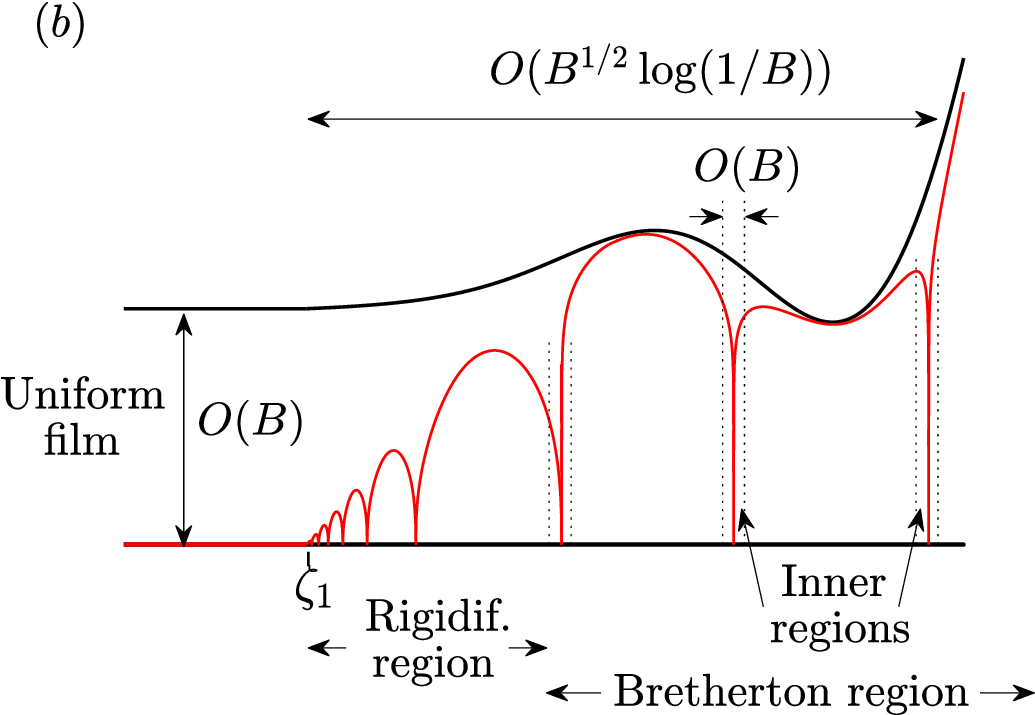}
    \end{subfigure}%
    \hfill
    \begin{subfigure}{0.49\textwidth}
        \centering
        \includegraphics[width = \textwidth]{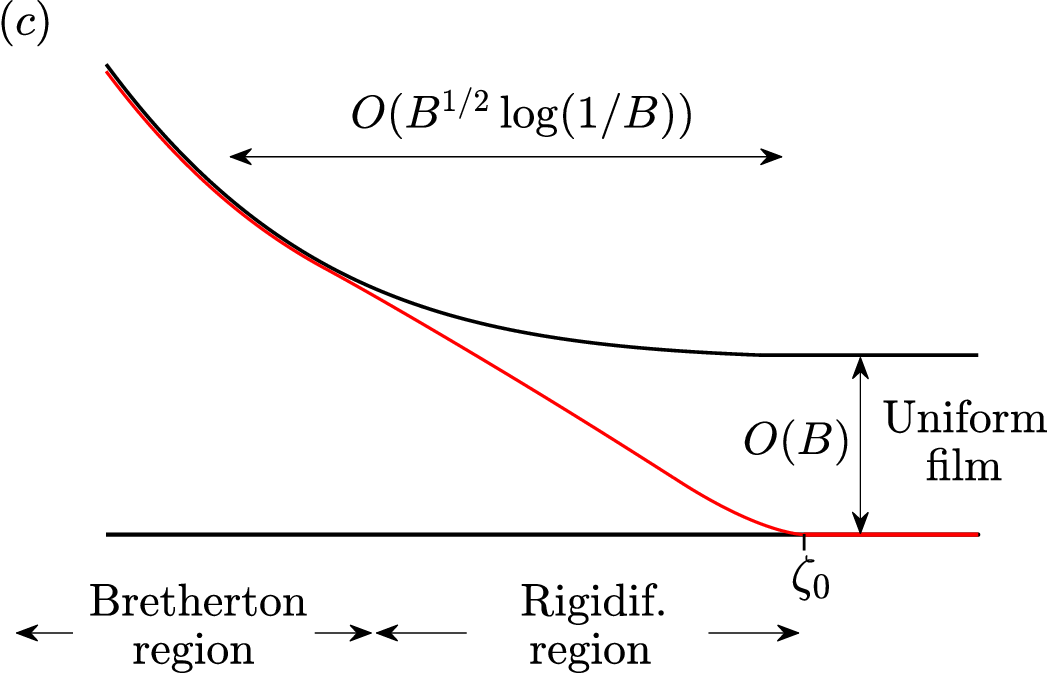}
    \end{subfigure}
    \caption{Sketch of the structure of the small-$B$ asymptotic solution. (a) View of the whole translating collar, showing $h$ (black), $Y$ (red), and the length scales of various asymptotic regions and the sizes of $h$ and $Y$ in several regions. (b) Close view of the front end of the collar, showing the Bretherton region, the unyielded uniform film ahead of the collar, the rigidification region where $Y\rightarrow0^+$, and the inner regions within the Bretherton region. (c) Close view of the rear end of the collar, showing the Bretherton region, the rigidification region and the unyielded uniform film behind the collar. }
    \label{fig:sketch_structure}
\end{figure*}

We expect the solution to \eqref{travelling_evoleqn}-\eqref{asymp_tauw} to be composed of a `main collar body', where $h=O(1)$ and where the interface shape is near to capillary-static equilibrium, and several short regions, at the two edges of the main body, where $h\ll1$, which enable the solution in the main body to match onto the unyielded uniform films ahead of and behind the collar (figure \ref{fig:sketch_structure}). Numerical solutions of \eqref{TFevoleqn} show that initially flat layers at relatively small $B$ reach such a state at late times (figure \ref{fig:nonlinear_evolution}a, $t=600$). In the main body of the collar, motivated by numerical simulations such as figure \ref{fig:nonlinear_evolution}(a), we assume that the fluid is yielded, so $Y>0$. Then, from \eqref{travelling_evoleqn} and \eqref{asymp_wp}, the leading-order solution for $h$ is the capillary-static shape satisfying
\begin{equation}
    h_\zeta+h_{\zeta\zeta\zeta} = 0,\label{asymp_hcapstateqn}
\end{equation}
\textcolor{black}{which is the same whether the fluid is Newtonian or viscoplastic.} We expect this capillary-static solution to be perfectly wetting on the tube wall, so we enforce boundary conditions, $h(0) = h_\zeta(0) = 0$, where we have chosen the front edge of the collar, which can be placed at any arbitrary point in $\zeta$, to lie at $\zeta=0$. Then, the capillary-static solution is a small-amplitude version of an unduloid,
\begin{equation}
    h \sim \frac{V}{2\pi}(1-\cos\zeta),\quad\mbox{for}\quad 0\leq \zeta\leq 2\pi\label{asymp_hcapstatsoln}
\end{equation}
where
\begin{equation}
    V = \int_0^{2\pi}h\,\mathrm{d}\zeta.
    \label{volume_constraint}
\end{equation}
The total volume of liquid in the collar is $2\pi V$, so the value of $V$ determines the collar volume. The length of the collar is forced to be $2\pi$ by \eqref{asymp_hcapstateqn}. We will confirm in \S\ref{sec:asymp_main_body}, by detailing higher-order corrections to the solution for $h$ in the collar body, that the leading-order solution \eqref{asymp_hcapstatsoln} is, indeed, consistent with $Y>0$ in the collar body.

In the remainder of this section, we detail how the capillary-static solution \eqref{asymp_hcapstatsoln} that forms the main body of the collar matches onto the uniform films ahead of and behind the collar via several short asymptotic regions at the edges of the collar. Given $B$, $\mathcal{J}$, $H^-_\infty$ and an initial value for $V$, we aim to determine the asymptotic structure of the solution from the leading film to the trailing film, and to find expressions for $\mathcal{U}$ and $H_\infty^+$. The volume of the collar can slowly change if $H_\infty^-\neq H_\infty^+$. From \eqref{travelling_evoleqn}, the flux takes a constant value, $q=-B^{5/2}\mathcal{U}H_\infty^-$, ahead of the collar, and another constant value, $q=-B^{5/2}\mathcal{U}H_\infty^+$, behind the collar. We can balance mass transfer into and out of the collar to determine how its volume gradually changes: integrating \eqref{travelling_evoleqn} along $0\leq \zeta\leq 2\pi$, and using \eqref{volume_constraint}, gives
\begin{equation}
    \frac{\mathrm{d}V}{\mathrm{d}\mathcal{T}} = \mathcal{U}\left( H_\infty^+ - H_\infty^- \right),
    \label{volume_change}
\end{equation}
assuming fluxes are uniform across the regions near $\zeta=0$ and $\zeta=2\pi$ where the film is very thin. Once we determine $\mathcal{U}$ and $H_\infty^+$, \eqref{volume_change} will describe the evolution of the collar's volume.

In the Newtonian problem \cite{jensen_draining_2000}, the solution for $h$ has a three-region structure: the main body of the collar is connected to two short `Bretherton' regions of length ${O}(B^{1/2})$, one at the front and one at the rear of the collar, which then match onto the films of uniform thickness ahead of, and behind, the translating body. We name these `Bretherton' regions given their similarity to analogous regions in the solution for the motion of long bubbles in liquid-filled tubes \cite{landau1942dragging,bretherton1961motion}. We will show that, in the limit $B\ll1$, similar asymptotic regions exist in the viscoplastic problem as in the Newtonian problem. However, multiple additional regions emerge, which must be considered in order to construct complete solutions for $h$ and $Y$. Figure \ref{fig:sketch_structure} illustrates the locations and sizes of the different asymptotic regions. Beside the Bretherton regions, in which the fluid turns out to be almost entirely yielded with $Y\approx h$ almost everywhere, there are `intermediate regions', with characteristic axial length scale of size $O(B^{1/4})$, between the main collar body and the Bretherton regions, in which $Y$ transitions from being close to $h$ to being small compared to $h$. There are also very short inner regions, of length $O(B)$, within the Bretherton region at the front of the collar (figure \ref{fig:sketch_structure}b), and there are `rigidification regions', which have characteristic length scale of size $O(B^{1/2})$, between the Bretherton region and the uniform film, in which the fluid transitions from being fully yielded to being unyielded; we leave detailed discussion of the inner regions within the front Bretherton region, and of the rigidification regions, to appendices \ref{sec:inbreth} and \ref{sec:rigidif}, respectively. Since the fluid in the uniform films ahead of and behind the collar must be unyielded, there must be two points, call them $\zeta=\zeta_0$ and $\zeta=\zeta_1$, such that $Y\rightarrow0^+$ as $\zeta\rightarrow \zeta_0^-$, $\zeta\rightarrow \zeta_1^+$, and $Y=0$ for $\zeta<\zeta_1$ or $\zeta>\zeta_0$. Since each of the regions at the edges of the main collar body is asymptotically short, we expect that $|\zeta_1|\ll1$ and $|\zeta_0-2\pi|\ll1$. The two rigidification regions are in neighbourhoods of the points $\zeta=\zeta_1$ and $\zeta=\zeta_0$, and we show in appendix \ref{sec:rigidif} below that the Bretherton regions sit at a distance of approximately $O(B^{1/2}\log(1/B))$ from $\zeta=\zeta_0$ and $\zeta=\zeta_1$, respectively (figure \ref{fig:sketch_structure}b,c). 

In the following subsection, we introduce the global force balance, which can be used to determine the translation speed of the collar. In the rest of \S\ref{sec:asymp_model}, we detail the solutions for $h$ and $Y$ in the various asymptotic regions, matching the solutions together, and finally arriving at expressions for $\mathcal{U}$ and $H_\infty^+$. 

\subsection{Global force balance}

As the collar translates, its weight is balanced by viscous shear stresses exerted on the fluid by the tube wall. We can integrate the wall shear stress \eqref{asymp_tauw} exerted on the fluid along the length of the collar, and further integration by parts gives
\begin{equation}
    \int_{\zeta_a}^{2\pi-\zeta_b}(-\tau_{\mathrm{w}})\,\mathrm{d}\zeta = \int_{\zeta_a}^{2\pi-\zeta_b}h(B-h_{\zeta}-h_{\zeta\zeta\zeta})\,\mathrm{d}\zeta \approx BV + \left[ \frac{1}{2}(h_\zeta)^2 - \frac{1}{2}h^2 - hh_{\zeta\zeta}\right]_{\zeta_a}^{2\pi-\zeta_b}.
    \label{globstressbal}
\end{equation}
Here, $\zeta_a$ and $\zeta_b$ are some constants such that $B^{1/2}\ll\zeta_a,\zeta_b\ll B^{1/4}$. The sizes of $\zeta_a$ and $\zeta_b$ ensure that stresses throughout the Bretherton regions, outside of the main collar body, are not included in this balance. From \eqref{volume_constraint}, the first term on the right hand side of \eqref{globstressbal} represents the weight of the collar, and the other terms represent contributions to the viscous drag from the front and rear edges of the collar, around $\zeta=0$ and $\zeta=2\pi$. If we can approximate each term in \eqref{globstressbal}, namely, the wall shear stress in $\zeta_a<\zeta<2\pi-\zeta_b$ (the left hand side of \eqref{globstressbal}) and the contributions to the viscous drag at the collar ends, then \eqref{globstressbal} can be used to determine an expression for the translation speed of the collar. This approach was used by Jensen \cite{jensen_draining_2000} for Newtonian collars, but we will show that there is a key difference in the force balance for viscoplastic collars: namely, the shear stress in the main collar body enters the leading-order force balance. This can be intuited since, for the fluid in the collar to be yielded, the stress must exceed the yield stress, which we have assumed is of the same order of magnitude as $B$, and, therefore, the same order of magnitude as the collar's weight.

\subsection{Main collar body}\label{sec:asymp_main_body}

In the main body of the collar, from \eqref{travelling_evoleqn}, we have $q_\zeta = O(B^{5/2})$, and we can assume that variations in flux occur over lengths comparable to that of the length of the collar, so $q = O(B^{5/2})$. We propose that the fluid in the collar's body is marginally yielded, and we expand as
\refstepcounter{equation}
$$
    h = h_0(\zeta) + B\log B h_1(\zeta) + Bh_2(\zeta) + \dots, \quad
    Y = B^{1/4}Y_0(\zeta) + B^{1/2}Y_1(\zeta) + \dots.
    \eqno{(\theequation{\mathit{a},\mathit{b}})}
    \label{body_hYexpansion}
$$
Given the expansions \eqref{body_hYexpansion}, \eqref{travelling_evoleqn} and \eqref{asymp_wp} imply
\refstepcounter{equation}
$$
    h_0' + h_0''' = 0,\quad 
    h_1' + h_1''' = 0,\quad 
    \frac{1}{2}(1-h_2'-h_2''')Y_0^2 + \mathcal{U} = 0,
    \eqno{(\theequation{\mathit{a},\mathit{b},\mathit{c}})}
    \label{body_heqns}
$$
and \eqref{YsmallBdefn} implies
\begin{equation}
     h_0(1-h_2'-h_2''') = \mathcal{J}.
    \label{body_Yeqn}
\end{equation}
In \eqref{body_Yeqn}, we have assumed that $1-h_2'-h_2'''\geq0$, which follows from (\ref{body_heqns}c) since we expect $\mathcal{U}\leq0$, and we have assumed that $Y>0$. 

The solution to (\ref{body_heqns}a) is the capillary-static solution \eqref{asymp_hcapstatsoln}, namely, 
\begin{equation}
    h_0 = \frac{V}{2\pi}(1-\cos\zeta).\label{body_h0}
\end{equation}
Since $h= O(B)$ away from the collar, we enforce boundary conditions on the first correction to $h$, 
\begin{equation}
    h_1(0) = h_1(2\pi) = 0,\label{body_h1BCs}
\end{equation}
and the volume constraint \eqref{volume_constraint} implies that 
\begin{equation}
    \int_0^{2\pi}h_1\,\mathrm{d}\zeta = \int_0^{2\pi}h_2\,\mathrm{d}\zeta = 0.
\label{body_volconstraint12}
\end{equation}
Given \eqref{body_h1BCs} and \eqref{body_volconstraint12}, we can then solve (\ref{body_heqns}b) to get
\begin{equation}
    h_1 = a_1\sin{\zeta},
    \label{body_h1}
\end{equation}
representing a small lateral displacement of the leading-order solution \eqref{body_h0}, and we can solve \eqref{body_Yeqn} to get
\begin{equation}
    h_2 = a_2\cos{\zeta} + b_2\sin{\zeta} - \pi(1-\cos{\zeta}) + \zeta +\frac{2\pi\mathcal{J}}{V}\left(-\zeta\cos{\zeta} + 2\sin{\zeta}\log\left[\sin\left(\frac{\zeta}{2}\right)\right]\right),
    \label{body_h2}
\end{equation}
for some constants $a_1$, $a_2$ and $b_2$ that will be determined below. The final three terms in \eqref{body_h2} capture a weak gravity-induced tilting of the collar. Finally, (\ref{body_heqns}c) and \eqref{body_Yeqn} give
\begin{equation}
    Y_0 = \sqrt{\frac{2|\mathcal{U}|h_0}{\mathcal{J}}} = \sqrt{\frac{|\mathcal{U}|V(1-\cos{\zeta})}{\pi\mathcal{J}}}.
    \label{body_Y0}
\end{equation}

The function $Y$ \eqref{YsmallBdefn} is defined such that the magnitude of the shear stress is $J=B\mathcal{J}$ at $y=Y$. Since, in the solution outlined above, we have $Y>0$ and $Y=O(B^{1/4})$ in the main collar body, we can deduce that, to leading order in $B$, $\tau_\mathrm{w}=J=B\mathcal{J}$ throughout the main body of the collar. Therefore, we have a leading-order approximation for the term on the left-hand-side of the global force balance \eqref{globstressbal},
\begin{equation}
    \int_{\zeta_a}^{2\pi-\zeta_b}(-\tau_\mathrm{w})\,\mathrm{d}\zeta \approx 2\pi B\mathcal{J}.
    \label{body_tauw}
\end{equation}

To aid the process of matching the solutions in the main body of the collar onto solutions in short regions at the collar's ends, we locally expand the solutions (\ref{body_h0}, \ref{body_h1}, \ref{body_h2}) for $h$ and \eqref{body_Y0} for $Y$. Near the front end of the collar,
\begin{equation}
    h \sim \frac{V}{4\pi}\zeta^2 + B\log{B}\,a_1\zeta + B\left(a_2 +\frac{4\pi\mathcal{J}}{V}\zeta\log{\zeta} + \zeta\left[b_2+1-\frac{2\pi\mathcal{J}}{V}(1+2\log{2})\right] \right) ~ \mbox{as}~ \zeta\rightarrow0^+,
    \label{body_hlim0}
\end{equation}
with errors of $O(\zeta^4,B\log{B}\zeta^3,B\zeta^2)$. Similarly,
\begin{equation}
    Y \sim B^{1/4}\zeta\sqrt{\frac{|\mathcal{U}|V}{2\pi\mathcal{J}}} \quad \mbox{as}\quad \zeta\rightarrow0^+,
    \label{body_Ylim0}
\end{equation}
with errors of $O(B^{1/4}\zeta^2,B^{1/2})$. Near the rear of the collar, 
\begin{multline}
    h\sim \frac{V}{4\pi}(\zeta-2\pi)^2 + B\log{B}\,a_1(\zeta-2\pi) + B\left( a_2 + 2\pi - \frac{4\pi^2\mathcal{J}}{V} + (\zeta-2\pi)\log{(2\pi-\zeta)}\frac{4\pi\mathcal{J}}{V}\right. \\\left.+ (\zeta-2\pi)\left[ b_2 + 1 -\frac{2\pi\mathcal{J}}{V}(1 +2\log{2})\right] \right) \quad\mbox{as}\quad \zeta\rightarrow 2\pi^-,
    \label{body_h_rear_limit}
\end{multline}
and
\begin{equation}
    Y \sim B^{1/4}(2\pi - \zeta)\sqrt{\frac{|\mathcal{U}|V}{2\pi\mathcal{J}}} \quad \mbox{as}\quad \zeta\rightarrow2\pi^-.
    \label{body_YlimL}
\end{equation}
The terms $B(2\pi - 4\pi^2\mathcal{J}/V)$ in \eqref{body_h_rear_limit}, absent in \eqref{body_hlim0}, reflect the weak gravity-induced tilting of the collar. They suggest that thickening of the trailing film is reduced by viscoplasticity as $\mathcal{J}$ increases from zero. 

\subsection{Intermediate regions}\label{sec:asymp_int}

As outlined above, the main collar body is marginally yielded with $Y\ll h$, but we will see below that the Bretherton regions are almost fully yielded with $Y\approx h$, implying an asymptotically thin pseudo-plug near the free surface. The intermediate regions that we detail in this section exist between the main body and the Bretherton regions, enabling a transition from a marginally-yielded state to a fully-yielded one.  

In these intermediate regions, we define local coordinates: at the front, we let $\zeta=B^{1/4}\hat{X}$, and at the rear, we let $\zeta=2\pi + B^{1/4}\hat{X}$. We propose expansions,
\begin{equation}
    h = B^{1/2}\frac{V}{4\pi}\hat{X}^2 + B\left(\hat{A}_\pm - \frac{V\hat{X}^4}{48\pi}\right) + B^{5/4}\hat{h}_\pm(\hat{X}) + \dots,\quad Y = B^{1/2}\hat{Y}_\pm(\hat{X})+\dots,
    \label{intermediate_expansions}
\end{equation}
where the subscripts `+' and `-' indicate quantities in the rear and front intermediate regions, respectively. The constants $\hat{A}_\pm$ will be determined shortly via matching. From \eqref{travelling_evoleqn}, we have that $q=O(B^{5/2})$, so inserting \eqref{intermediate_expansions} into \eqref{travelling_evoleqn} gives
\begin{equation}
    \frac{1}{2}\hat{Y}_\pm^2\hat{h}_\pm''' = \frac{3\mathcal{U}V\hat{X}^2}{3V\hat{X}^2 - 4\pi\hat{Y}_\pm}.
    \label{intermediate_evoleqn}
\end{equation}
Combining \eqref{intermediate_evoleqn} with the yield stress condition \eqref{YsmallBdefn} gives
\begin{equation}
    -6\mathcal{U}V\hat{X}^2\left(\frac{V\hat{X}^2}{4\pi} - \hat{Y}_\pm\right)
     = \mathcal{J}\hat{Y}^2_\pm(3V\hat{X}^2 - 4\pi\hat{Y}_\pm).
     \label{intermediate_Yeqn}
\end{equation}
The cubic equation \eqref{intermediate_Yeqn} can be solved to find $\hat{Y}_\pm(\hat{X})$ and \eqref{intermediate_evoleqn} can then be integrated to find $\hat{h}_\pm$. In the far field approaching the main collar body, we expect $Y\ll h$, so $\hat{Y}_\pm\ll\hat{X}^2$ as $\hat{X}\rightarrow\mp\infty$. Therefore, \eqref{intermediate_Yeqn} implies
\begin{equation}
    \hat{Y}_\pm\sim \sqrt{\frac{|\mathcal{U}|V}{2\pi\mathcal{J}}}|\hat{X}|\quad\mbox{as}\quad \hat{X}\rightarrow\mp\infty,
    \label{intermediate_Ylims}
\end{equation}
and we assume
\begin{equation}
    \hat{h}_\pm\sim \frac{4\pi\mathcal{J}}{V}\hat{X}\log|\hat{X}| 
    \quad\mbox{as}\quad\hat{X}\rightarrow\mp\infty,
    \label{int_h'''inflim}
\end{equation}
which is consistent with \eqref{intermediate_evoleqn}. 

In order to match the intermediate region solutions \eqref{intermediate_expansions} onto the main body solution \eqref{body_hlim0} and \eqref{body_h_rear_limit}, we write the intermediate region solution in terms of $\zeta$. At the front, where $\zeta\ll1$,
\begin{equation}
    h\sim \frac{V}{2\pi}\left(\frac{\zeta^2}{2} - \frac{\zeta^4}{24}\right) - B\log{B}\frac{\pi\mathcal{J}}{V}\zeta + B\left[\hat{A}_- + \frac{4\pi\mathcal{J}}{V}\zeta\log \zeta\right]+\dots,
    \label{intermediate_matching_front}
\end{equation}
which, after matching with \eqref{body_hlim0}, implies $a_1 = -\pi\mathcal{J}/V$, $a_2 = \hat{A}_-$ and $b_2 = -1+2\pi\mathcal{J}(1+2\log{2})/V$. We will determine $\hat{A}_-$ below by matching with the Bretherton region. At the rear, where $|\zeta-2\pi|\ll1$, \eqref{intermediate_Yeqn} and \eqref{intermediate_evoleqn} imply
\begin{equation}
    h\sim \frac{V}{2\pi}\left(\frac{(\zeta-2\pi)^2}{2} - \frac{(\zeta-2\pi)^4}{24}\right) - B\log{B}\frac{\pi\mathcal{J}}{V}(\zeta-2\pi) + B\left[\hat{A}_+ + \frac{4\pi\mathcal{J}}{V}(\zeta-2\pi)\log(2\pi - \zeta)\right].
    \label{intermediate_matching_rear}
\end{equation}
Matching \eqref{intermediate_matching_rear} and \eqref{body_h_rear_limit} gives 
\begin{equation}
    \hat{A}_+ = \hat{A}_- + 2\pi - 4\pi^2\mathcal{J}/V.\label{Ahat+Ahat-Match}
\end{equation}
The expansion \eqref{intermediate_Ylims} for $Y$ is consistent with both \eqref{body_Ylim0} and \eqref{body_YlimL}.

\subsection{Bretherton regions}\label{sec:breth}

In the Bretherton regions, we also define local coordinates: at the front, we let $\zeta = B^{1/2}\xi$, and at the rear, we let $\zeta = 2\pi + B^{1/2}\xi$. In these regions, we expect that the fluid is almost entirely yielded with only asymptotically small pseudo-plugs at the free surface, and that the leading-order problem resembles the classical Newtonian analogue. Therefore, we expand as
\refstepcounter{equation}
$$
    h = B\mathcal{H}_\pm(\xi) + \dots, \quad Y = B\mathcal{H}_\pm(\xi)+ \dots,
    \eqno{(\theequation{\mathit{a},\mathit{b}})}
    \label{Breth_expansions}
$$
where the subscripts again distinguish quantities in the rear and front Bretherton regions. The evolution equation \eqref{travelling_evoleqn} again says that the leading-order flux is constant, with $q = -B^{5/2}\mathcal{U}H_\infty^+$ in the rear Bretherton region, and $q = -B^{5/2}\mathcal{U}H_\infty^-$ in the front. Therefore, \eqref{travelling_evoleqn} and \eqref{Breth_expansions} give
\begin{equation}
    \frac{1}{3}\mathcal{H}_\pm^3\mathcal{H}_{\pm,\xi\xi\xi} = \mathcal{U}(\mathcal{H}_\pm - H_\infty^\pm),
    \label{bretherton_eqn1}
\end{equation}
which is the classical equation from the Newtonian problem \cite{bretherton1961motion,jensen_draining_2000}. 

We introduce rescaled variables, defining $X_\pm = (3|\mathcal{U}|)^{1/3}\xi/{H_\infty^{\pm}}$, and let $\mathcal{H}_\pm=H_\infty^\pm H_\pm(X_\pm)$, so that \eqref{bretherton_eqn1} becomes
\begin{equation}
    H_\pm^3H_\pm''' = 1 - H_\pm.
    \label{bretherton_eqn_rescaled}
\end{equation}
At the rear, we expect that $H_+\rightarrow1$ as $X_+\rightarrow\infty$, and at the front that $H_-\rightarrow1$ as $X_-\rightarrow-\infty$. Each of these leading-order problems then has the same solution as in the Newtonian problem \cite{wong1995motion}: there is a unique solution for $H_+$ such that
\begin{equation}
    H_+ \sim \frac{1}{2}H_2X_+^2 + H_0 + \frac{2}{3H_2^2X_+} + \dots \quad\mbox{as}\quad X_+\rightarrow-\infty,
    \label{Breth_H+}
\end{equation}
where $H_2\approx0.6430$ and $H_0\approx2.8996$, and the solutions for $H_-$ have the far-field behaviour,
\begin{equation}
    H_-\sim \frac{1}{2}G_2X_-^2 + G_0 + \dots \quad\mbox{as}\quad X_-\rightarrow\infty,
    \label{Breth_H-}
\end{equation}
where $G_2$ and $G_0$ are constants, and $G_0$ is determined by $G_2$. If $G_2 = H_2$, then $G_0\approx-0.8453$ \cite{wong1995motion}. Equation \eqref{bretherton_eqn_rescaled} suggests that both $H_+$ and $H_-$ decay exponentially towards $1$ in the far field away from the collar, with $H_+$ decreasing monotonically towards $1$ as $X_+\rightarrow\infty$, and $H_-$ undulating, repeatedly passing through $H_-=1$, as $X_-\rightarrow-\infty$ \cite{kalliadasis1994drop,jensen_draining_2000}. However, in each case, the description provided by \eqref{bretherton_eqn_rescaled} must break down at some finite $\xi$, where the fully-yielded state transitions to an unyielded state. We will discuss the transition from the Bretherton region to the unyielded uniform film in more detail appendix \ref{sec:rigidif}.

To match onto the intermediate region solutions, we expand the Bretherton region solutions in terms of the outer variable $\hat{X}$. At the rear, from \eqref{Breth_H+},
\begin{equation}
    h \sim B^{1/2}H_2\hat{X}^2\frac{\left(3|\mathcal{U}|\right)^{2/3}}{2H_\infty^+} + BH_\infty^+ H_0 + B^{5/4}\frac{2(H_\infty^+)^2}{3H_2^2(3|\mathcal{U}|)^{1/3}\hat{X}}+\dots\label{breth+match}
\end{equation}
Expanding the intermediate solution \eqref{intermediate_expansions} for $\hat{X}\ll1$ gives
\begin{equation}
    h \sim B^{1/2}\frac{V}{4\pi}\hat{X}^2 + B\hat{A}_+ - B^{5/4}\frac{8\pi^2\mathcal{U}}{3V^2\hat{X}} + \dots.\label{int+match}
\end{equation}
Matching \eqref{breth+match} and \eqref{int+match} gives 
\begin{equation}
    VH_\infty^+ = 2\pi H_2(3|\mathcal{U}|)^{2/3},
    \label{MatchingCondVHinf+}
\end{equation}
and $\hat{A}_+ = H_\infty^+ H_0$. Similarly, at the front, 
\begin{equation}
    h \sim B^{1/2}G_2\hat{X}^2\frac{\left(3|\mathcal{U}|\right)^{2/3}}{2H_\infty^-} + BH_\infty^- G_0 +\dots\label{breth-match}
\end{equation}
and the intermediate region solution \eqref{intermediate_expansions}, using \eqref{Ahat+Ahat-Match}, for $\hat{X}\ll1$, gives
\begin{equation}
    h \sim B^{1/2}\frac{V}{4\pi}\hat{X}^2 + B\left(\hat{A}_+ - 2\pi+\frac{4\pi^2
    \mathcal{J}}{V}\right)- B^{5/4}\frac{8\pi^2\mathcal{U}}{3V^2\hat{X}} + \dots.\label{int-match}
\end{equation}
Matching \eqref{breth-match} and \eqref{int-match} gives 
\begin{equation}
    \frac{G_2}{H_\infty^-} = \frac{V}{2\pi(3|\mathcal{U}|)^{2/3}} = \frac{H_2}{H_\infty^+},
    \label{G2H2reln}
\end{equation}
and 
\begin{equation}
    \hat{A}_+ = H_\infty^+ H_0 = H_\infty^-G_0 + 2\pi - \frac{4\pi^2\mathcal{J}}{V}.
    \label{breth_Ahat+}
\end{equation}

Combining \eqref{G2H2reln} and \eqref{breth_Ahat+}, we can write
\begin{equation}
    F(\mu) = \frac{2\pi}{H_\infty^-}\left(1 - \frac{2\pi\mathcal{J}}{V}\right) \quad\mbox{where}\quad \mu = \frac{H_\infty^+}{H_\infty^-} \quad\mbox{and}\quad F(\mu)\equiv \frac{1}{G_2}\left(H_0H_2 - G_0G_2\right).
    \label{breth_Fdefn}
\end{equation}
Jensen \cite{jensen_draining_2000} showed that 
\begin{equation}
    F(\mu)\approx3.745 + 5.02(\mu-1)
    \label{breth_Fapprox}
\end{equation}
is a good approximation for $\tfrac{1}{2}\leq \mu\leq\tfrac{3}{2}$. The collar speed and trailing film thickness can be deduced from \eqref{G2H2reln} and \eqref{breth_Fdefn}, and so the evolution of the collar volume can then be determined via \eqref{volume_change}. The analogous results for Newtonian collars \cite{jensen_draining_2000} are recovered by setting $\mathcal{J}=0$ in \eqref{breth_Fdefn}. 

We can recover \eqref{breth_Fdefn} without using \eqref{breth_Ahat+}, which was derived from matching of higher-order terms, by appealing to the global force balance \eqref{globstressbal}. We can use the limiting forms of the Bretherton region solutions, \eqref{Breth_H+} and \eqref{Breth_H-}, to approximate the last three terms in \eqref{globstressbal}. From \eqref{Breth_H+} and \eqref{Breth_H-}, we find that
\begin{equation}
    \left[-hh_{\zeta\zeta}\right]^{2\pi-\zeta_b}_{\zeta_a}\approx B(3|\mathcal{U}|)^{2/3}\left(G_0G_2 - H_0H_2\right),
    \label{breth_hhzz}
\end{equation}
since $\zeta_a, \zeta_b\gg B^{1/2}$, and that the remaining two terms in the square brackets in \eqref{globstressbal} are smaller than $O(B)$. We interpret \eqref{breth_hhzz} as the contribution to the global force balance from the viscous drag in the two Bretherton regions, which are imbalanced because of the small differences between the front and rear of the collar. The same approximation \eqref{breth_hhzz} could equally be reached by calculating $hh_{\zeta\zeta}$ from the inner limits \eqref{intermediate_matching_front} and \eqref{intermediate_matching_rear} of the intermediate region solutions, given the values of $\hat{A}_\pm$ from matching with the Bretherton region solutions. Inserting \eqref{breth_hhzz} and \eqref{body_tauw} into \eqref{globstressbal} gives the leading-order force balance across the collar,
\begin{equation}
    2\pi B\mathcal{J} = BV + B(3|\mathcal{U}|)^{2/3}\left(G_0G_2 - H_0H_2\right),
    \label{globalstressbalanceMATCHED}
\end{equation}
which, when combined with the result \eqref{G2H2reln} derived from leading-order matching, exactly recovers \eqref{breth_Fdefn}. 

We have now derived a leading-order expression for $\mathcal{U}$, and relations between $H_\infty^\pm$, $\mathcal{U}$, $\mathcal{J}$ and $V$. There are further details of the asymptotic model, discussed in appendices \ref{sec:inbreth} and \ref{sec:rigidif}, which are required to expose the full structure of the solution. In the front Bretherton region, the layer height, $\mathcal{H}$, has an undulating shape, and there is a sequence of points within the Bretherton region around which $\mathcal{H}'''\approx0$. An asymptotically short inner region, in which $Y$ is no longer equal to $h$ at leading order, must exist around each point where $\mathcal{H}'''=0$. We discuss the structure of these inner regions in appendix \ref{sec:inbreth}. At the outer edges of both Bretherton regions, $Y$ must approach zero in order for the solution to match onto the uniform unyielded film away from the collar. In appendix \ref{sec:rigidif}, we discuss the structure of two `rigidification regions' near to the points $\zeta=\zeta_0$ and $\zeta=\zeta_1$, where $Y\rightarrow0$. We find that in the rigidification regions, $\zeta-\zeta_1=O(B^{1/2})$ and $\zeta_0-\zeta=O(B^{1/2})$, respectively, but with $\zeta-\zeta_1\ll B^{1/2}\log(1/B)$ and $\zeta_0-\zeta\ll B^{1/2}\log(1/B)$, and that the Bretherton region solutions outlined above strictly hold when $\zeta-\zeta_1\gg B^{1/2}\log(1/B)$ and $\zeta_0-\zeta\gg B^{1/2}\log(1/B)$, respectively. This motivates the $O(B^{1/2}\log(1/B))$ length scales identified in figure \ref{fig:sketch_structure}(b,c) for the distance between the Bretherton regions and the points $\zeta=\zeta_0$ and $\zeta=\zeta_1$. \textcolor{black}{The rigidification regions resemble analogous regions close to yield surfaces in other thin-film capillary flows \cite{jalaal_long_2016,jalaal_stoeber_balmforth_2021}, but in the current problem, the rigidification regions are always matched onto almost fully-yielded Bretherton regions. Appendix \ref{sec:rigidif} outlines how the rigidification regions connect the Bretherton regions to the unyielded uniform films away from the collar. } None of these details impact the leading-order predictions for $\mathcal{U}$ and $H_\infty^+$ determined above.

\section{Translating collars at small $B$}\label{sec:results_smallB}

The asymptotic model outlined in \S\ref{sec:asymp_model} provides information on how a single, isolated collar propagates into a uniform film ahead of it when $B\ll1$ and $\mathcal{J}=O(1)$. Numerical solutions of \eqref{TFevoleqn}, such as the one shown in figure \ref{fig:nonlinear_evolution}, can also describe the motion of a translating collar when they are run to sufficiently long times to allow for a collar to develop and then begin steadily propagating. Since we enforce periodicity at the sides of our computational domain when solving \eqref{TFevoleqn} numerically, the film thicknesses at the front edge and rear edge of a translating collar must be the same. This means that the numerical simulations always reach a state at late times in which the collar is steadily translating. Therefore, we can directly compare predictions from the asymptotic model, in the case that $V$ and $\mathcal{U}$ are constant and $H_\infty^+=H_\infty^-$, against late-time predictions from numerical simulations: we do so in \S\ref{sec:results_steady}. In \S\ref{sec:results_unsteady}, we discuss how an isolated collar evolves in an unsteady manner when the film thickness ahead and behind are unequal, which can be described by the asymptotic model.

\subsection{Steady translation}\label{sec:results_steady}

We consider a single, isolated collar of volume $2\pi V$, translating at speed $\mathcal{U}$, which can be described by the asymptotic model from \S\ref{sec:asymp_model}. We enforce $H_\infty^+ = H_\infty^-$, so that the volume of fluid entering the collar at the front edge is equal to the volume exiting at the rear edge. Therefore, $V$ does not change with time and the translation of the collar is steady. In this case, \eqref{breth_Fdefn} gives the necessary relationship between the film thickness and the volume,
\begin{equation}
    \frac{H_\infty^-F(1)V}{2\pi} = V - 2\pi\mathcal{J},
    \label{res:steady_HinfV}
\end{equation}
where $F(1) \approx3.745$. Combining \eqref{G2H2reln} with \eqref{res:steady_HinfV} gives the asymptotic prediction for the steady propagation speed,
\begin{equation}
    (3|\mathcal{U}|)^{2/3} = \frac{V - 2\pi\mathcal{J}}{H_2(H_0-G_0)},
    \label{res:steady_U}
\end{equation}
where $H_2\approx0.6430$, $H_0\approx2.8996$ and $G_0\approx-0.8453$. 

For steady translation to be possible, the forces on the collar must balance. We can interpret \eqref{res:steady_HinfV} as a statement of the force balance: the term on the left-hand-side is the contribution from the viscous drag at the ends of the collar, the first term on the right is from the collar's weight, and the second term on the right is from the viscous drag across the body of the collar. Since the viscous drag at the ends and the weight are both proportional to $V$, forces cannot balance when $\mathcal{J}=0$, except at one specific value of the film thickness ahead of the collar, $H_\infty^-=2\pi/F(1)$. When $\mathcal{J}=0$, the value of $V$ has no impact on the force balance \eqref{res:steady_HinfV}; $V$ just sets the collar speed via \eqref{res:steady_U}. By contrast, the viscous drag across the collar body is independent of $V$, so when $\mathcal{J}>0$, changing $V$ does affect the balance of forces on the collar, which provides a mechanism for steady solutions to exist for more values of $H_\infty^-$ compared to the Newtonian case. When $\mathcal{J}>0$, a steadily-translating collar may exist for any value of $H_\infty^-$ in the range $0<H_\infty^-\leq 2\pi/F(1)$, as long as $V$ and $H_\infty^-$ satisfy \eqref{res:steady_HinfV}. 

 \begin{figure}
     \centering
     \includegraphics[width=\linewidth]{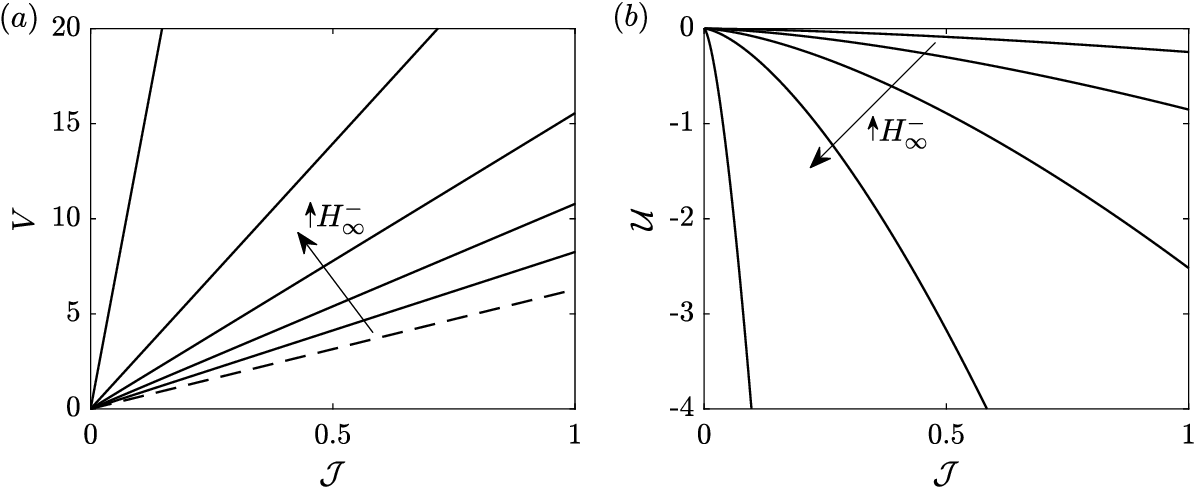}
     \caption{Dependence of collar volume and speed on Bingham number, $\mathcal{J}$, for fixed values of $H_\infty^-$. Variation of (a) $V$, and (b) $\mathcal{U}$, for $0<\mathcal{J}<1$, according to \eqref{res:steady_HinfV} and \eqref{res:steady_U}. Each solid line corresponds to one value of $H_\infty^-$, with the values shown being $H_\infty^-=\{0.4,0.7,1,1.3,1.6\}$. The large arrows indicate increasing values of $H_\infty^-$. The dashed line in (a) is $V=2\pi\mathcal{J}$, which corresponds to the minimum volume required for the collar to translate. 
     }
     \label{fig:UVvsJfixedH}
 \end{figure}

If we consider the film thickness ahead of the collar, $H_\infty^-$, to be fixed, then the value of $V$ required for steady motion to occur is determined via \eqref{res:steady_HinfV} and \eqref{res:steady_U}, assuming that $0<H_\infty^-\leq2\pi/F(1)$. In this case, the effect of increasing $\mathcal{J}$ is to increase both the volume and speed of steadily-translating collars (figure \ref{fig:UVvsJfixedH}). To understand this result, we again appeal to the balance of forces required for steady motion, described by \eqref{res:steady_HinfV}: if $\mathcal{J}$ is increased, then the size of the drag force on the collar increases, so for the force balance to remain satisfied while $H_\infty^-$ remains fixed, the weight of the collar, or equivalently its volume, must increase. The speed of the collar increases proportionally to $V^{3/2}$ when $H_\infty^-$ is fixed, so increasing $\mathcal{J}$ also leads to an increase in the steady collar speed. In \S\ref{sec:results_unsteady} below, we discuss how a collar translating into a film with fixed thickness ahead of it may gradually adjust its volume to reach a steadily-translating state.

\begin{figure}
    \centering
    \includegraphics[width=\linewidth]{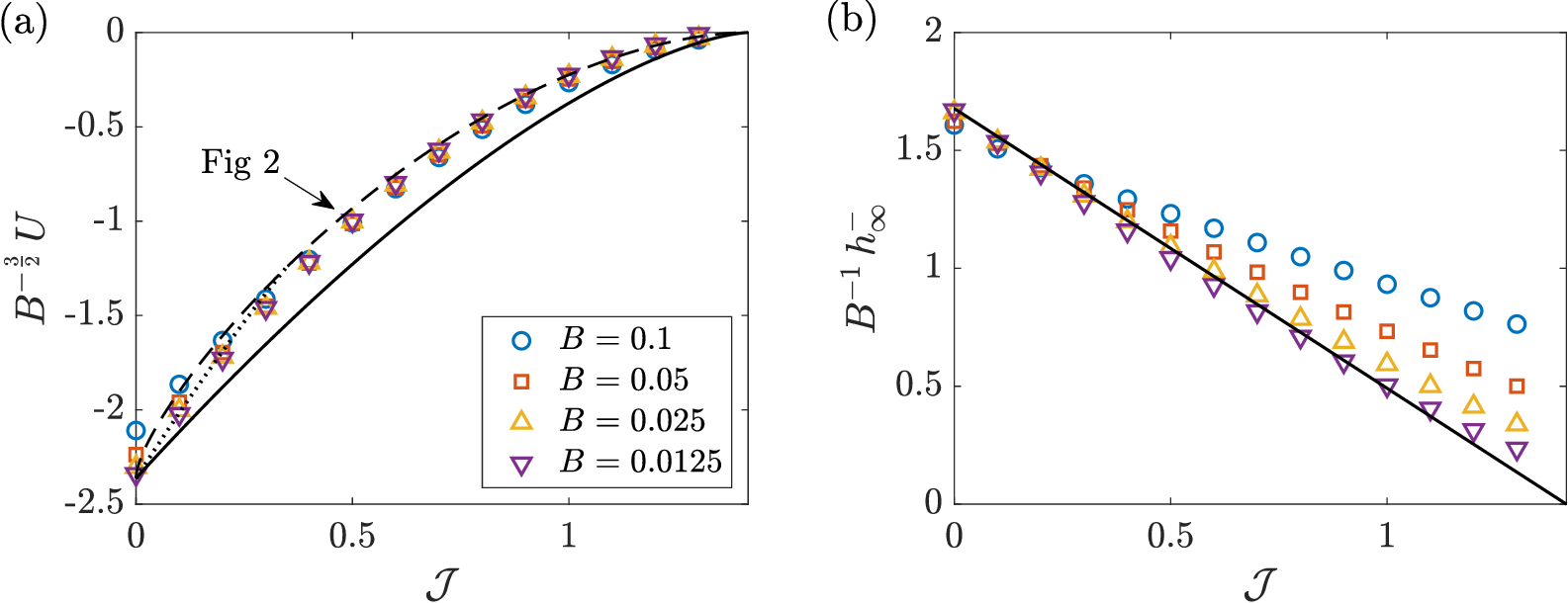}
    \caption{\textcolor{black}{Data from numerical solutions of \eqref{TFevoleqn} with domain length $\mathcal{L}=2\pi/k_\mathrm{m}$, for various values of the Bond number, $B$, and Bingham number, $\mathcal{J}=J/B$. (a) Late-time collar speed, $\mathcal{U}=B^{-3/2}U$, evaluated as the gradient of $z_{\max}(t)$ at the end of the simulation, where $z_{\max}$ is the location of the maximum in $h$. (b) $B^{-1}h(z_{\max}\pm \mathcal{L}/2,t_{\mathrm{end}})$, which approximates the deposited film thickness, $H_\infty^-=B^{-1}h_\infty^-$, where $t_{\mathrm{end}}$ is the simulation end time. Solid lines show leading-order asymptotic predictions \eqref{res:steady_U} for $\mathcal{U}$ and \eqref{res:steady_HinfV} for $H_\infty^-$, evaluated by setting $V=\mathcal{L}$. The dashed line in (a) is $\mathcal{U}$ evaluated using \eqref{app_GSBcomp}, the global force balance that includes $O(B^{5/4}\log{B})$ corrections to the wall shear stress, for $B=0.025$. The dotted line shows $\mathcal{U}$ determined via \eqref{GSB_J32} and \eqref{tildeF89}, the global force balance in the case that $J=O(B^{3/2})$, again for $B=0.025$. For each $B$, $t_{\mathrm{end}}$ is chosen to be large enough to achieve adequate convergence, ranging from $t_{\mathrm{end}}=8\times10^3$ for $B=0.1$ to $t_{\mathrm{end}}=4\times10^4$ for $B=0.0125$. The initial conditions \eqref{hIC} are used for simulations with $\mathcal{J}=0$. For each $B$, $\mathcal{J}$ is then increased in increments of $0.1$ and the final solution from the previous value of $\mathcal{J}$ is used as the initial conditions; this enables faster convergence to the late-time solutions.} }
    \label{fig:UvsJ}
\end{figure}

Rather than treating $H_\infty^-$ as fixed, as we do in figure \ref{fig:UVvsJfixedH}, we could instead treat $V$ as fixed, considering a collar with fixed volume.
In this case, from \eqref{res:steady_HinfV} and \eqref{res:steady_U}, both the film thickness, $H_\infty^-$, and the propagation speed, $|\mathcal{U}|$, decrease as $\mathcal{J}$ is increased. The case of fixed $V$ is more amenable to direct comparison with numerical solutions of \eqref{TFevoleqn} in periodic domains, since the total volume of fluid within the periodic domain is fixed to be $\mathcal{L}$. 
The asymptotic prediction for $\mathcal{U}$ compares well with late-time collar speeds in numerical simulations (figure \ref{fig:UvsJ}\textcolor{black}{a}, solid line), \textcolor{black}{as does the prediction for $H_\infty^-$ when $B$ is sufficiently small (figure \ref{fig:UvsJ}b).} In figure \ref{fig:UvsJ}\textcolor{black}{(a)}, there is good (but not perfect) agreement between $B^{-3/2}U$ from numerical simulations, and $\mathcal{U}$, which provides evidence to support the scalings and solutions in the asymptotic theory for small $B$. Even when $B$ is not extremely small, there is reasonably good quantitative agreement between the numerical and asymptotic predictions for translation speed (figure \ref{fig:UvsJ}), although the asymptotics generally predict a modestly faster speed. At smaller values of $\mathcal{J}$, the speeds from numerical solutions appear to converge towards the asymptotic prediction as $B$ is decreased, but at larger $\mathcal{J}$ there is not clear convergence towards the leading-order asymptotic prediction for the values of $B$ tested in figure \ref{fig:UvsJ}. When $\mathcal{J}$ is large enough that viscoplastic effects are significant, we may expect that convergence to the small-$B$ asymptotic regime only occurs at much smaller $B$ than when $\mathcal{J}\approx0$. This is because the largest error term in $Y$ in the main body of the collar is only expected to be $O(B^{1/4})$ larger than the leading-order term \eqref{body_hYexpansion}, so $B$ must be very small before the error in $Y$, and by extension the error in the viscous drag in the collar, is negligible. 

\textcolor{black}{Figure \ref{fig:UvsJ}(b) shows a comparison between $H_\infty^-$ from \eqref{res:steady_HinfV} and $B^{-1}h(z_\mathrm{max}\pm\mathcal{L}/2,t_\mathrm{end})$, which is an approximation to the deposited film thickness at the end time of numerical simulations. There is a more prominent discrepancy in the deposited film thickness when $B$ is not small for larger values of $\mathcal{J}$, but the agreement generally improves as $B$ is decreased. This suggests that higher-order terms in the small-$B$ expansion for $H_\infty^+$ may increase in magnitude as $\mathcal{J}$ is increased, meaning the error between leading-order asymptotics and numerics is larger at larger $\mathcal{J}$. Nonetheless, when $B=0.0125$, there is good agreement between numerics and the leading-order approximation for $H_\infty^+$ for all values of $\mathcal{J}$ tested in figure \ref{fig:UvsJ}. }

\textcolor{black}{For numerical simulations with $\mathcal{J}>1$, there would not be unstable growth and collar formation if the initial conditions \eqref{numIC} were used. However, there is still a subcritical instability that can be triggered if a sufficiently large perturbation is applied to the free surface initially. The asymptotic solution exists for $0\leq\mathcal{J}<V/2\pi$, and steadily-translating collar states are achieved at late-times in simulations throughout this range (figure \ref{fig:UvsJ}). For $\mathcal{J}>V/2\pi$, we find that there can still be unstable growth in numerical simulations if a large perturbation is applied to the free surface initially (data not shown), but that collars generally appear to decelerate at late times, with the collar speed approaching zero, rather than reaching a steadily-translating state. In the limit $\mathcal{J}\rightarrow (V/2\pi)^-$, we have $\mathcal{U}\rightarrow0^-$, so the assumption that the collar speed scales like $U\sim B^{-3/2}$ could break down when $\mathcal{J}$ is sufficiently close to $V/2\pi$. An alternative scaling may be required to fully explore this limit, and a modified asymptotic solution could exist. However, this limit represents only extremely slow collar motion, so we do not explore this in detail, and continue to focus on the results where $\mathcal{U}=O(1)$. }


Improved approximations for the steady collar speed may be found by determining a higher-order correction to the wall shear stress, $\tau_{\mathrm{w}}$, and inserting it into the global force balance \eqref{globstressbal}; details are given in appendix \ref{sec:appHOcorrections}. We find that the viscous drag on the collar is approximately given by
\begin{equation}
    \int(-\tau_{\mathrm{w}})\,\mathrm{d}\zeta \sim 2\pi\mathcal{J}B + B^{5/4}\log\left(\frac{1}{B}\right)\sqrt{\frac{2\pi|\mathcal{U}|\mathcal{J}}{V}}+\dots,
    \label{tauwHO}
\end{equation}
where the $O(B^{5/4}\log(1/B))$ correction arises from contributions from the wall shear stress in the intermediate regions. Replacing the left-hand side of \eqref{globalstressbalanceMATCHED} with the expression \eqref{tauwHO} yields a prediction for the collar speed which generally agrees more closely with numerical results than the leading-order prediction for $\mathcal{U}$ (figure \ref{fig:UvsJ}\textcolor{black}{a}, dashed line). 

For very weak viscoplasticity, one expects to recover a quasi-Newtonian collar in which $Y\approx h$. The solution structure illustrated in figure \ref{fig:sketch_structure} for $J=O(B)$ with $B\ll1$ has $Y\ll h$ in the collar. This suggests the existence of an intermediate regime, with $0<J\ll B$, for which $Y$ and $h$ are comparable in the collar. In appendix \ref{sec:appJB32}, we derive the resulting approximate expression for the collar speed in the case that $J=B^{3/2}\hat{J}$, with $\hat{J}=O(1)$. We find that
\begin{equation}
    BV\approx B(3|\mathcal{U}|)^{2/3}(H_0H_2-G_0G_2) + \tilde{F}(\Gamma)B^{3/2}\hat{J}
    \label{GSB_J32},
\end{equation}
where $\Gamma\equiv\hat{J}V/|\mathcal{U}|$, and we find that
\begin{equation}
    \tilde{F}(\Gamma) \approx 8.9\label{tildeF89}
\end{equation}
is a good approximation for $0<\Gamma<20$. Numerical evaluation of $\tilde{F}(\Gamma)$ suggests that $\tilde{F}\approx 2\pi$ when $\Gamma$ is sufficiently large, suggesting that the result \eqref{globalstressbalanceMATCHED} for $J=O(B)$ can be recovered in the limit $\hat{J}\rightarrow\infty$. The collar speed determined via \eqref{GSB_J32}, with the approximation \eqref{tildeF89}, yields an improved estimate of the steady collar speed when $\mathcal{J}\ll1$ (figure \ref{fig:UvsJ}\textcolor{black}{a}, dotted line), compared to the leading-order prediction \eqref{globalstressbalanceMATCHED} for $\mathcal{U}$. For $B=0.025$, $\Gamma<20$ holds for $0<\mathcal{J}\leq0.4$, so use of \eqref{tildeF89} is justified at least for $0<\mathcal{J}\leq0.4$ in figure \ref{fig:UvsJ}\textcolor{black}{(a)}.

\begin{figure}
    \centering
    \includegraphics[width=\linewidth]{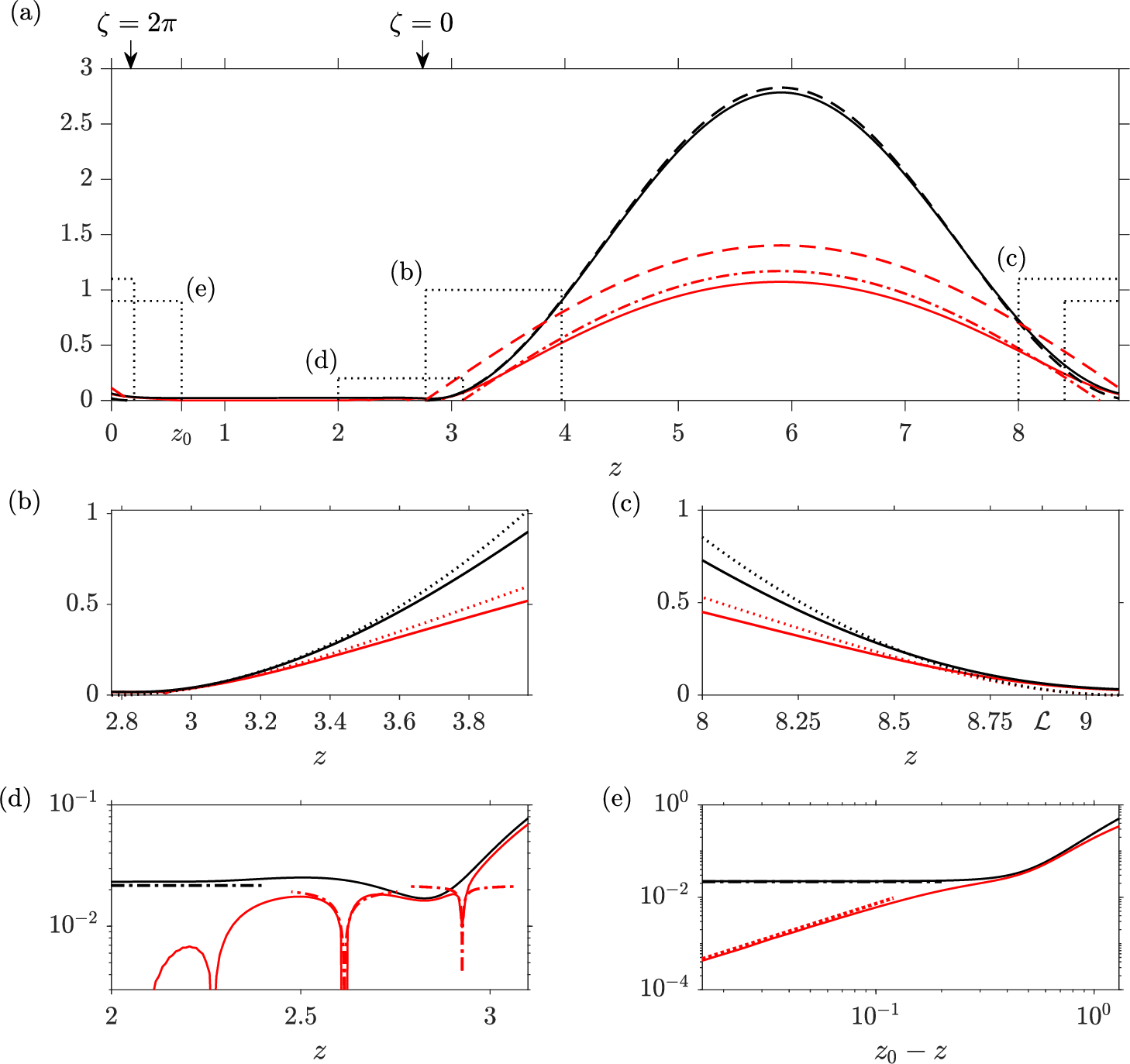}
    \caption{Numerical solution of \eqref{TFevoleqn}, with $\mathcal{J}=0.5$ and $B=0.02$, at $t=3000$, showing $h$ (solid black) and $Y$ (solid red). (a) Dashed lines show leading-order asymptotic predictions \eqref{body_h0} and \eqref{body_Y0}. 
    First-order prediction for $Y$ \eqref{Yfirstorder} also shown (dot-dashed). (b) Front intermediate region. Leading-order predictions from \eqref{intermediate_expansions} and \eqref{intermediate_Yeqn} are shown for $h$ (dotted black) and $Y$ (dotted red). (c) Rear intermediate region, as in (b). (d) Front edge of the collar. Asymptotic predictions are shown for the uniform film thickness \eqref{res:steady_HinfV} (dot-dashed black) where we have used the approximation $V=\mathcal{L}$ to calculate $H_\infty^-$, and for inner limits in the Bretherton region \eqref{inbreth_Y} (dot-dashed red), where we use the gradient in $h$ from the numerical solution to approximate $\alpha_n$ in \eqref{inbreth_Y}. (e) Rear edge, where $z_0=0.62$ is approximately where $Y\rightarrow0^+$ in the numerical solution. Asymptotic predictions are shown for the film thickness \eqref{res:steady_HinfV} (dot-dashed black) and the inner limit \eqref{rigidif_innerexpansions} (dotted red).}
    \label{fig:structure_comparison}
\end{figure}

As well as the giving the relationships \eqref{res:steady_HinfV} and \eqref{res:steady_U} between the collar volume, film thickness and propagation speed, the leading-order asymptotic model also provides approximations for the detailed structure of $h$ and $Y$. Figure \ref{fig:structure_comparison} shows comparison of a late-time solution from numerics against predictions from the asymptotic model after rescaling the solutions in the various asymptotic regions, given the finite value of $B=0.02$. In the asymptotic solution in figure \ref{fig:structure_comparison}, we used a fixed value for the collar volume, $V=\mathcal{L}$, equal to the total volume of fluid in the periodic domain from the numerical simulation. We find good qualitative agreement, and often quantitative agreement, between the structure of $h$ and $Y$ at late times in the numerical solution and the predictions from the asymptotic model. The leading-order asymptotic prediction for $h$ in the main collar body \eqref{body_h0} agrees well with a numerical solution at $B=0.02$ (figure \ref{fig:structure_comparison}a). The error in the leading-order prediction for $Y$ \eqref{body_Y0} is larger, although there is still qualitative agreement in the shape (figure \ref{fig:structure_comparison}a). In appendix \ref{sec:appHOcorrections}, we derive a correction to the leading-order expression for $Y$ in the main collar body,
\begin{equation}
    Y \sim B^{1/4}\sqrt{\frac{2|\mathcal{U}|h_0}{\mathcal{J}}} - B^{1/2}\frac{2|\mathcal{U}|}{3\mathcal{J}} + \dots,
    \label{Yfirstorder}
\end{equation}
which yields improved agreement with numerical results (figure \ref{fig:structure_comparison}a). 

Figures \ref{fig:structure_comparison}(b) and \ref{fig:structure_comparison}(c) show that the leading-order intermediate-region solutions, which are unique to the viscoplastic problem, provide good agreement with $h$ and $Y$ at the front and rear ends of the main collar body. The prediction \eqref{res:steady_HinfV} for the uniform film thickness agrees closely with the film thickness in the numerical solution (figure \ref{fig:structure_comparison}d,e). The structure of $Y$ in two of the inner regions in the front Bretherton region are shown in figure \ref{fig:structure_comparison}(d): there is good agreement in one inner region, centred around $z\approx2.6$, which supports the general structure derived in appendix \ref{sec:inbreth} for these regions, although the agreement with the asymptotics in the inner region centred around $z\approx2.9$ is weaker. The error in the latter inner region is largely caused by the fact that the slope of $h$ is relatively large at that point, whilst the leading-order asymptotic solution assumes $h\approx BH_\infty^-$. The finite resolution of the numerical finite-difference scheme inevitably restricts our ability to capture all of the infinite sequence of undulations predicted in appendix \ref{sec:rigidif}, so we cannot easily test any further predictions for $Y$ in the front rigidification region. However, we find that in the rear rigidification region (figure \ref{fig:structure_comparison}e), there is close agreement between the numerical solution close to the point where $Y\rightarrow0^+$ and the inner limit predicted from asymptotics \eqref{rigidif_innerexpansions}.


\subsection{Unsteady evolution}\label{sec:results_unsteady}

We now consider a single, isolated collar translating into a film with a fixed thickness $BH_\infty^-$, but no longer assume that the motion is steady, nor that $H_\infty^-=H_\infty^+$. We use \eqref{volume_change} to describe the slow changes in the volume of the collar that occur when $H_\infty^-\neq H_\infty^+$. When the film thicknesses ahead of, and behind, the collar are similar, but not equal, we can combine \eqref{G2H2reln}, \eqref{breth_Fdefn} and \eqref{breth_Fapprox} to give an approximate expression for the translation speed when $J=O(B)$,
\begin{equation}
    (3|\mathcal{U}|)^{2/3} \approx \frac{VH_\infty^-}{2\pi H_2}\left[1 + \frac{1}{F'(1)}\left(\frac{2\pi}{H_\infty^-}-F(1)\right)\right] - \frac{2\pi\mathcal{J}}{H_2F'(1)},
    \label{UlinearF}
\end{equation}
where $F(1)$ and $F'(1)$ are given in \eqref{breth_Fapprox}. Similarly, we can combine \eqref{volume_change} with \eqref{breth_Fdefn} and \eqref{breth_Fapprox} to reach
\begin{equation}
    \frac{\mathrm{d}V}{\mathrm{d}\mathcal{T}} \approx \frac{|\mathcal{U}|H_\infty^-}{F'(1)}\left[F(1) - \frac{2\pi}{H_\infty^-}\left(1 - \frac{2\pi\mathcal{J}}{V}\right) \right],
    \label{dVdTlinearF}
\end{equation}
which predicts the slow evolution of the collar volume in terms of the leading film thickness, $H_\infty^-$, the speed, $|\mathcal{U}|$, and the Bingham number, $\mathcal{J}$. The collar grows if the quantity in the square bracket in \eqref{dVdTlinearF} is positive, and shrinks if that quantity is negative. 

\begin{figure}
    \centering
    \includegraphics[width=\linewidth]{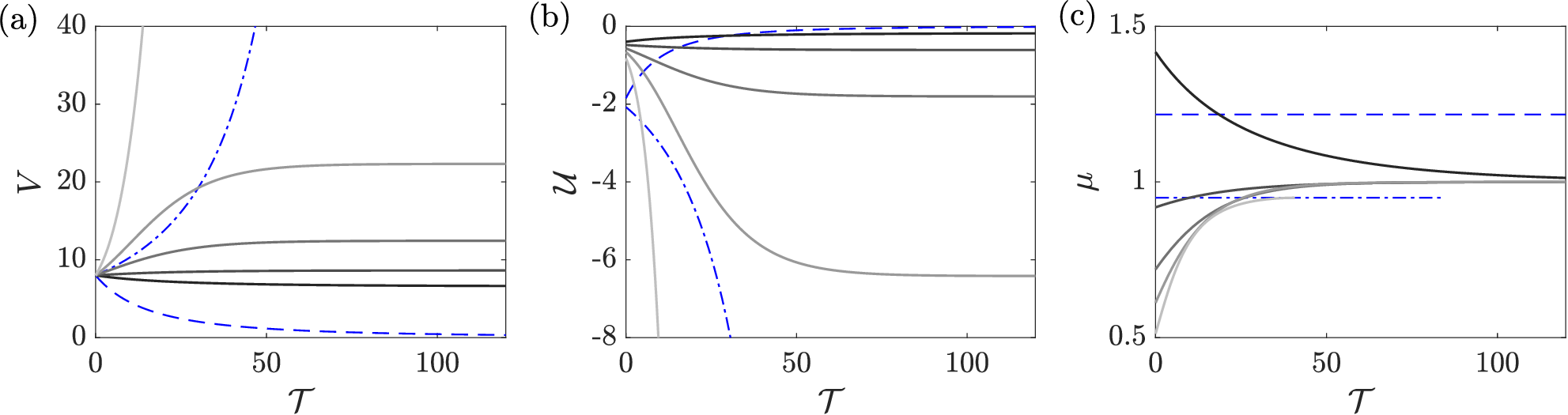}
    \caption{Time evolution of collars according to \eqref{UlinearF} and \eqref{dVdTlinearF}, showing (a) $V$, where the collar volume is $2\pi V$, (b) $\mathcal{U}$, the collar speed, and (c) $\mu=H_\infty^+/H_\infty^-$, the trailing film thickness relative to the film thickness ahead of the collar, versus time. Five solutions are shown with $\mathcal{J}=0.8$, initial volume given by $V(0)=8$, and various film thicknesses, $H_\infty^-=\{0.4,0.7,1,1.3,1.8\}$. Darker lines indicate smaller values of $H_\infty^-$. Two solutions are shown for $\mathcal{J}=0$, with $V(0)=8$ and film thicknesses $H_\infty^-=1.3$ (dashed blue) and $H_\infty^-=1.8$ (dot-dashed blue). }
    \label{fig:dVdT1}
\end{figure}

When $\mathcal{J}>0$, there are two possible outcomes, depending on the film thickness, $H_\infty^-$, and the initial value of $V$: either the collar approaches a steadily-translating state, or the collar's volume blows up. If $H_\infty^- \geq 2\pi/F(1)$, then $V$ increases subsequently for all time, regardless of the current value of $V$, and so the volume blows up. As blow-up occurs, the final term in the square bracket in \eqref{dVdTlinearF} becomes small, and the last term in \eqref{UlinearF} becomes much smaller than the other terms, so the collar behaves like a Newtonian collar near blow-up, with $V$ increasing proportional to $(\mathcal{T}_0-\mathcal{T})^{-2}$, where $\mathcal{T}_0$ is some blow-up time. If $H_\infty^- < 2\pi/F(1)$, then the collar approaches the steadily-translating solution in which $V=V_s\equiv4\pi^2\mathcal{J}/(2\pi - F(1)H_\infty^-)$. This can be seen from \eqref{dVdTlinearF}: if $2\pi\mathcal{J}<V<V_s$ then $\mathrm{d}V/\mathrm{d}\mathcal{T}>0$ and if $V>V_s$ then $\mathrm{d}V/\mathrm{d}\mathcal{T}<0$. When $V\approx V_s$, \eqref{dVdTlinearF} indicates that $V$ approaches $V_s$ exponentially quickly. 
Figure \ref{fig:dVdT1} shows examples of the evolution of collar volume, according to \eqref{dVdTlinearF} and \eqref{UlinearF}, for $\mathcal{J}=0.8$. When $H_\infty^-$ is sufficiently small, $V$ can decrease from its initial value before reaching steady state; when $H_\infty^-$ is larger, but less than $2\pi/F(1)$, $V$ can increase significantly before reaching steady state. When $H_\infty^-> 2\pi/F(1)$ we observe blow-up in $V$. 

There are some clear contrasts between the quasi-steady evolution of viscoplastic collars and that of Newtonian collars. Setting $\mathcal{J}=0$, \eqref{dVdTlinearF} then describes how the volume of a Newtonian collar changes in time: if $H_\infty^->2\pi/F(1)$ then there is finite-time blow up, as in the viscoplastic case, but if $H_\infty^-<2\pi/F(1)$ then the volume decreases for all subsequent time until $V\rightarrow0$ (see the examples in figure \ref{fig:dVdT1}). As we noted in \S\ref{sec:results_steady}, steadily-translating Newtonian collars can exist only if $H_\infty^-=2\pi/F(1)$, but \eqref{dVdTlinearF} indicates that this state is unstable: if $H_\infty^-$ takes any other value, the collar either blows up or shrinks and disappears. By contrast, when $\mathcal{J}>0$, for any film thickness in the range $0<H_\infty^- < 2\pi/F(1)$, the volume of the collar adjusts to reach a steadily-translating state. Hence, introducing a yield stress stabilises the system for a significant range of film thicknesses for which there are no stable, steadily-translating states in the Newtonian case. 


\section{Discussion}\label{sec:discussion}

We have investigated the motion and evolution of thin films of viscoplastic fluid coating vertical cylindrical tubes. We derived an evolution equation for the dimensionless film thickness, $h$, using thin-film theory, before constructing asymptotic solutions for quasi-steadily translating collars of fluid in the limit of small Bond number, $B\ll1$. These solutions exhibit an intricate structure, with multiple asymptotic regions that must be matched together. The structure shares some common features with the Newtonian analogue \cite{jensen_draining_2000}: the near-capillary-static main body of the collar, and the short `Bretherton' regions at the front and rear edges of the collar, the viscous drag from which regulates the collar translation speed. However, several features are novel in the viscoplastic problem: intermediate regions emerge between the main collar body and the Bretherton regions that are required to make $Y$ (the internal boundary between regions of shear-dominated and plug-like flow) continuous; inner regions emerge within the front Bretherton region, where $Y$ becomes small, and `rigidification' regions emerge between the Bretherton regions and the uniform film, in which $Y\rightarrow0^+$. Moreover, the global force balance is altered when a yield stress is introduced as the viscous drag in the main body of the collar enters the leading-order balance, which then modifies the expression for the collar speed. 

We solved the thin-film evolution equation numerically to examine the nonlinear evolution of an initially nearly-flat layer. Numerical simulations were conducted in periodic domains, which allowed for investigation of the initial growth of the instability, as well as the subsequent development of a steadily-translating collar. We investigated predictions from the small-$B$ model in the case that there is one steadily-translating collar with a fixed volume, and compared the results to numerical solutions of the full thin-film evolution equation. We found generally good agreement between the structure of the asymptotic model in its various regions with the shapes of $h$ and $Y$ in a numerical simulation at $B=0.02$ (figure \ref{fig:structure_comparison}). We also found generally good agreement between leading-order asymptotic predictions for the collar speed and the corresponding quantity from numerical simulations for a range of $\mathcal{J}$, even when $B$ was not extremely small (figure \ref{fig:UvsJ}\textcolor{black}{a}). Including higher-order corrections in the global force balance, arising from contributions from the intermediate regions, resulted in improved predictions for the collar speed. We also used the asymptotic model to show that, when the film thickness ahead of the collar is fixed, increasing $\mathcal{J}$ leads to an increase in both the volume and speed of a steadily-translating collar (figure \ref{fig:UVvsJfixedH}).

There are notable differences in the existence and stability of steadily-translating states in the viscoplastic problem compared to the Newtonian problem. In the Newtonian problem there is exactly one film thickness ahead of the collar for which a steadily-translating collar can exist, which is unstable in the sense that any small change in the film thickness will cause the collar volume to either shrink to zero or blow up in finite time. By contrast, when there is a yield stress, we identified a range of film thicknesses for which steadily-translating states exist, and showed that, for these film thicknesses, the volume of a collar adjusts exponentially quickly to a value at which a steadily-translating state is reached. This stabilisation of collar propagation by viscoplasticity can be explained by considering the forces acting on the collar, which must balance for steady propagation to occur. The collar's weight and the viscous drag at the edges of the collar are both proportional to the collar's volume, whilst the viscous drag across the collar's body is proportional to the Bingham number, $\mathcal{J}$, but not the volume. Therefore, for a Newtonian collar, its volume has no effect on this force balance, but for a viscoplastic collar, forces can balance given the right collar volume. Steady translation of a viscoplastic collar is achieved when gradual changes in the collar's volume cause the net force on the collar to decrease towards zero. 

Our results may have implications for understanding how mucus drains under gravity through small airways in the lungs, where the Bond number may be small but gravitational effects may still drive complex dynamics. Airway mucus is known to have a yield stress, which is typically small in healthy lungs but increases in diseases such as cystic fibrosis or COPD \cite{patarin_rheological_2020,kavishvar2023yielding}. Our quantification of how increasing the liquid yield stress can slow or halt instability and draining of a film coating a tube, may reflect a mechanism for decreased gravitational drainage of mucus in diseased airways, or for how increased drainage my be triggered by application of drugs, such as mucolytics, that decrease mucus yield stress \cite{patarin_rheological_2020}. However, the applicability of our work to modelling airway mucus has limitations that remain to be addressed before the model could be considered physiologically complete, including the effects of: airway wall elasticity; rheological properties such as elasticity, shear-thinning or thixotropy; presence of surfactant; shear stress from air flow; or thickening of the mucus layer which may occur in disease. 

Extending our model to describe motion of thick collars or liquid plugs, as has been done for the Newtonian case \cite{jensen_draining_2000}, is left for future studies. 
\textcolor{black}{Computational fluid dynamics (CFD) simulations have been used to study the problem of pressure-driven liquid plug motion \cite{zamankhan_steady_2012}, which exhibits some features similar to those of gravity-driven collar motion, including an undulating free surface at the leading edge of the plug, with the shear-rate and shear stress rapidly changing signs where these undulations occur. However, direct comparison of our theoretical results with CFD is left for future studies in which gravity-driven collar motion is simulated.} The finite-time blow-up of slender translating collars that can occur when the film ahead of the collar is sufficiently thick (figure \ref{fig:dVdT1}) suggests that, in these cases, collars do not remain slender and finite-thickness effects must then be taken into account. It is likely that, where the thin-film theory predicts blow-up, either the collar will saturate as a non-slender collar, will form a liquid plug in the tube, or, if it is on the exterior of the tube, may form a large droplet. In future, finite-thickness effects could be investigated by deriving and analysing a long-wave model, of the type used to predict plug formation in the Rayleigh-Plateau instability previously \cite{shemilt2022surface,shemilt2023surfactant,ogrosky_linear_2021,camassa_2014_gravity,camassa_2017_air,camassa_viscous_2015,halpern_effect_2010}, or by solving the full two-dimensional equations via CFD, which has also been used to study plug formation for Newtonian and non-Newtonian fluids. \cite{romano_liquid_2019,romano_effect_2021,erken_2022_elastoviscoplastic,fazla2024effects} 

There has been some experimental work characterising viscoplastic effects on plug propagation and rupture \cite{Bahrani_propagation_2022}, against which a theoretical model of plug propagation could be compared. \textcolor{black}{Bahrani et al. \cite{Bahrani_propagation_2022} also developed a model of elastoviscoplastic plug propagation, which took into account the thickening of the liquid film deposited behind the plug due to elastic effects. Employing a more complex constitutive law in our model may modify its predictions for the collar dynamics, but this is beyond the scope of the current study. } It has been shown that plug formation and propagation can induce significant stresses on the tube wall, which may have physiological importance since airway wall epithelial cells can be injured if large enough stresses are exerted on them \cite{huh_acoustically_2007}. In our asymptotic model of quasi-steadily translating collars, a stress approximately equal to the liquid's yield stress is exerted on the tube wall across the length of the collar. This suggests that taking into account the liquid's yield stress may be important when considering wall stresses exerted by collar or plug propagation. However, an extended model that can model plug propagation and non-slender collars would be required to confirm this. 

We have focused on modelling the formation and motion of individual liquid collars. In numerical simulations, we have used periodic domains with \textcolor{black}{a fixed length that allows for only a single collar to develop in the domain}. In longer domains, multiple collars may develop and potentially interact; we leave investigation of the dynamics of interacting viscoplastic collars for future studies. 

Whilst we derived our models assuming the liquid layer coats the interior of a cylinder, our modelling is also applicable to thin films coating the exterior of a cylinder, so there are additional potential applications to our work in the context of dip-coating of fibres with viscoplastic fluids. Experimental studies have been conducted in which rods are dip-coated in viscoplastic fluid \cite{smit2019stress,smit2021withdrawal}, but Rayleigh-Plateau instability has not been reported, perhaps because liquid yield stresses or cylinder diameters were too large in these experiments. If the low-$B$, low-$J$ regime that we have focused on were accessed experimentally, the formation and motion of viscoplastic liquid collars could be investigated in a dip-coating experiment and our theoretical predictions could be tested. 


We have described how viscoplastic films on vertical cylindrical tubes can form liquid collars, and have characterised how these collars propagate. We have highlighted key differences between the viscoplastic and Newtonian problems, notably in the asymptotic structure of quasi-steadily translating collar solutions, in the speed of collar propagation, and in the stability of steadily-translating collar solutions for various film thicknesses ahead of the oncoming collar. Weak gravitational effects, when the Bond number is low, can drive complex dynamics, and we have shown how a similarly small yield stress can modify those dynamics significantly.

\begin{acknowledgments}
J.D.S. was supported by an EPSRC doctoral training award. J.D.S, A.H. and C.A.W. are supported by the ESPRC Network Plus `BIOREME' (EP/W000490/1). C.A.W., A.H. and O.E.J. are supported
by the National Institute for Health and Care Research (NIHR) Manchester Biomedical Research Centre (BRC)
(NIHR203308). The views expressed in this publication are those of the author(s) and not necessarily those of the NHS, the National Institute for Health Research, Health Education England or the Department of Health.
\end{acknowledgments}

\appendix

\section{Small-$B$ asymptotics in a low-yield-stress regime}\label{sec:appJB32}

\setcounter{equation}{0}
\renewcommand\theequation{A.\arabic{equation}}

In the asymptotic model in \S\ref{sec:asymp_model}, we assumed that the capillary Bingham number was of the same order of magnitude as the Bond number, i.e., $J = O(B)$. Here, we investigate a regime in which $J$ is asymptotically smaller than $B$, where viscoplasticity may still influence the dynamics, although perhaps to a lesser extent. We let $J = B^{3/2}\hat{J}$, with $\hat{J}=O(1)$. In the main collar body, we then expand as
\begin{equation}
    h = \hat{h}_0 + B\hat{h}_1 + B^{3/2}\hat{h}_2 + \dots, \quad Y = \hat{Y}_0 + \dots, \quad \tau_\mathrm{w} = B^{3/2}\hat{\tau}_0+\dots,
\end{equation}
allowing the yield surface to lie in the middle of the collar. Then \eqref{travelling_evoleqn} implies that
\begin{equation}
    \hat{h}_0 = h_0 = \frac{V}{2\pi}(1-\cos\zeta),\label{appB_h0}
\end{equation}
\begin{equation}
    \hat{h}_1 = A_1\cos\zeta + A_2\sin\zeta + \zeta,
\end{equation}
for some constants $A_1$ and $A_2$. The leading-order wall shear stress is given by
\begin{equation}
    \hat{\tau}_0 = h_0(\hat{h}_{2,\zeta}+\hat{h}_{2,\zeta\zeta\zeta}) = \frac{2\mathcal{U}h_0^2}{\hat{Y}_0^2(h_0-\hat{Y}_0/3)},
    \label{appB_tau0}
\end{equation}
and \eqref{YsmallBdefn} implies that
\begin{equation}
    \hat{Y}_0 = h_0\left(1 - \frac{\hat{J}}{|\hat{\tau}_0|}\right).
    \label{appB_Y0}
\end{equation}
Combining \eqref{appB_tau0} and \eqref{appB_Y0} gives a cubic equation for $Y_0$,
\begin{equation}
    \frac{\hat{J}}{3}\hat{Y}_0^3 - \hat{J}h_0\hat{Y}_0^2 + 2h_0\mathcal{U}(\hat{Y}_0-h_0) = 0. 
    \label{appB_Y0cubic}
\end{equation}
We expect that the Bretherton regions are classical, to leading order at least, and that the collar propagation speed can be determined to leading order by matching $\hat{h}_1$ onto the Bretherton region solutions. Since $\hat{h}_1$ has the same structure as in the Newtonian case, we also expect that the leading-order prediction for the collar speed will be the same as for a Newtonian collar. We can, however, approximate the first correction to the collar speed induced by viscoplasticity by determining the first correction to the viscous drag exerted at the wall, and find an expression for the collar speed via a balance between the viscous drag and the collar's weight, as in \eqref{globstressbal}. 

Given that the Bretherton regions are classical to leading order, the integral of the wall shear stress across the whole collar (including the Bretherton regions) is composed of leading-order contributions from the Bretherton regions and a higher-order correction from the main body of the collar. We write the integral of the wall shear stress across the whole collar's length, including the first correction, as
\begin{equation}
    \int(-\tau_\mathrm{w})\,\mathrm{d}\zeta 
    \sim B(3|\mathcal{U}|)^{2/3}\left(I_- + I_+\right) -
    B^{3/2}\int_\beta^{2\pi-\beta}\hat{\tau}_0\,\mathrm{d}\zeta +\dots,
    \label{app_B32_tauwintegral}
\end{equation}
where
\begin{equation}
    I_- = \int_{-\infty}^{\frac{(3|\mathcal{U}|)^{1/3}}{B^{1/2}H_\infty^-}\beta}(-H_-H_-''')\,\mathrm{d}X_-
    \quad\mbox{and}\quad 
    I_+ = \int^{\infty}_{\frac{-(3|\mathcal{U}|)^{1/3}}{B^{1/2}H_\infty^+}\beta}(-H_+H_+''')\,\mathrm{d}X_+
    \label{Ipmdefn}
\end{equation}
are the contributions from the Bretherton regions, with the functions $H_\pm(X_\pm)$ defined as in \S\ref{sec:breth} above, and $\beta$ is some constant such that $B^{1/2}\ll \beta \ll 1$. Integrating by parts in \eqref{Ipmdefn} and using \eqref{breth+match} and \eqref{breth-match}, the expansions of $H_\pm$ in the respective limits $X_\pm\rightarrow\mp\infty$, we have that
\begin{equation}
    I_-\sim -G_0G_2 - B^{1/2}\frac{4\pi(3|\mathcal{U}|)^{1/3}}{\beta V},\quad\mbox{and}\quad I_+\sim H_0H_2 - B^{1/2}\frac{4\pi(3|\mathcal{U}|)^{1/3}}{\beta V}.
    \label{Ipm}
\end{equation}
Inserting \eqref{Ipm} into \eqref{app_B32_tauwintegral}, we have
\begin{equation}
    \int(-\tau_\mathrm{w})\,\mathrm{d}\zeta \sim 
    B(3|\mathcal{U}|)^{2/3}(H_0H_2-G_0G_2) - B^{3/2}\mathcal{U}I_0,
\end{equation}
where
\begin{equation}
    I_0=\lim_{\beta\to0}\left[ -\frac{24\pi}{\beta V} +\int_\beta^{2\pi-\beta}\frac{2h_0^2}{\hat{Y}_0^2(h_0-\hat{Y}_0/3)}\,\mathrm{d}\zeta \right].
    \label{I0defn}
\end{equation}
In arriving at \eqref{I0defn}, we have used \eqref{appB_tau0}. Exploiting the symmetry of the integrand in \eqref{I0defn} about $\zeta=\pi$, and removing the singularity in the integrand at $\zeta=0$, we can write
\begin{equation}
    I_0 = \lim_{\beta\to0}\left[ \frac{1}{V}\int_\beta^{\pi}\left(\frac{4\bar{h}_0^2}{\bar{Y}_0^2(\bar{h}_0-\bar{Y}_0/3)} -\frac{24\pi}{\zeta^2}\right)\,\mathrm{d}\zeta -\frac{24}{V} \right],\label{I0}
\end{equation}
where $\bar{h}_0=h_0/V$ and $\bar{Y}_0=\hat{Y}_0/V$. 

\begin{figure}
    \centering
    \includegraphics[width=\linewidth]{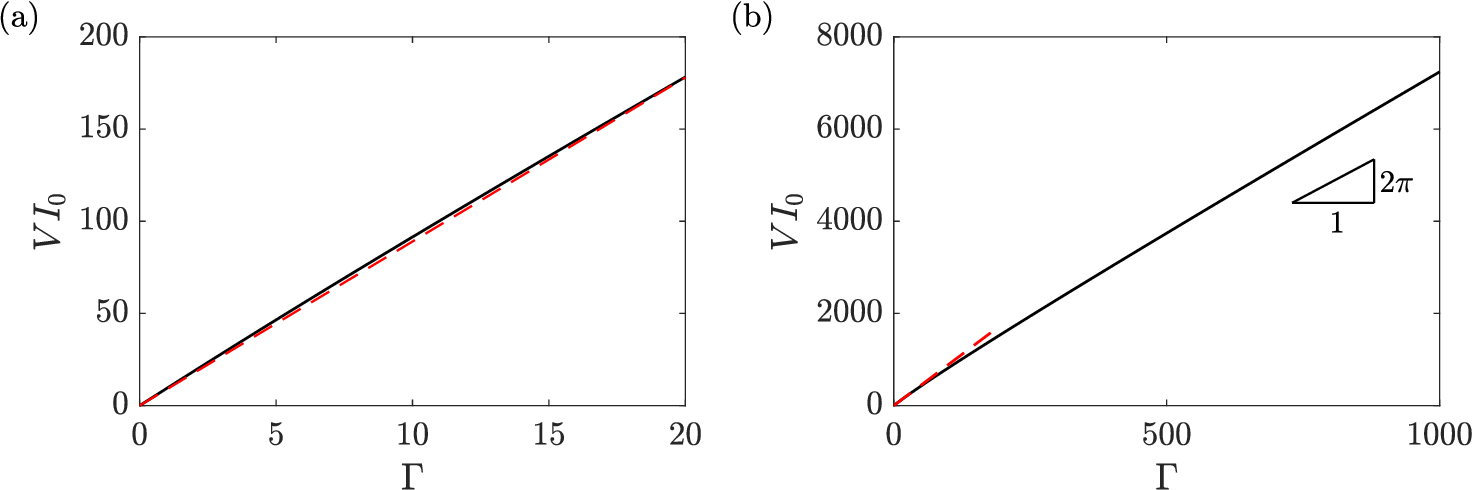}
    \caption{Solution to \eqref{I0} with \eqref{appB_Ybar0cubic} and \eqref{appB_h0} (black). Red dashed lines show the linear approximation, $VI_0\approx8.9\Gamma$. (a) $VI_0$ is approximated well by $8.9\Gamma$ for $0<\Gamma<20$. (b) At large values of $\Gamma$, the slope of $VI_0$ approaches $2\pi$. }
    \label{fig:appB_VI0}
\end{figure}

After rewriting \eqref{appB_Y0cubic} as
\begin{equation}
    \Gamma\bar{Y}_0^2\left(\bar{h}_0 - \frac{1}{3}\bar{Y}_0\right) + 2\bar{h}_0(\bar{Y}_0-\bar{h}_0)=0 \quad \mbox{where}\quad \Gamma \equiv \frac{\hat{J}V}{|\mathcal{U}|},
    \label{appB_Ybar0cubic}
\end{equation}
we can evaluate $VI_0$, via \eqref{I0}, \eqref{appB_Ybar0cubic} and \eqref{appB_h0}, numerically for a given value of $\Gamma$ (figure \ref{fig:appB_VI0}). Then, noting that the viscous drag \eqref{app_B32_tauwintegral} must balance with the collar's weight, we have
\begin{equation}
     BV \approx B(3|\mathcal{U}|)^{2/3}(H_0H_2-G_0G_2) + B^{3/2}|\mathcal{U}|I_0.
     \label{appB_GSB}
\end{equation}
A good approximation for $I_0$ is given by $VI_0\approx8.9\Gamma$, at least when $0<\Gamma<20$ (figure \ref{fig:appB_VI0}a). For 
$B\geq0.025$, $\Gamma<20$ holds for $0\leq\mathcal{J}\leq0.4$. Substituting the linear approximation, $I_0\approx8.9\Gamma$, into \eqref{appB_GSB}, we find
\begin{equation}
    BV \approx B(3|\mathcal{U}|)^{2/3}(H_0H_2-G_0G_2) + 8.9B^{3/2}\hat{J}.
    \label{appB32_GSBlinapprox}
\end{equation}
The rightmost term in \eqref{appB32_GSBlinapprox} approximates the first correction to the global force balance due to viscoplasticity, which has the same form as in \eqref{globalstressbalanceMATCHED}, but with a different prefactor: 8.9 in \eqref{appB32_GSBlinapprox} instead of $2\pi$ in \eqref{globalstressbalanceMATCHED}. As $\Gamma$ is increased, the slope of $VI_0(\Gamma)$ decreases from $8.9$ and approaches $2\pi$ at large $\Gamma$ (figure \ref{fig:appB_VI0}b), so the result \eqref{globalstressbalanceMATCHED} can be recovered, indicating that the theory in this appendix in the limit of large $\hat{J}$ is consistent with the theory in \S\ref{sec:asymp_model}, as expected. 

\begin{figure}
    \centering
    \includegraphics[width=0.7\linewidth]{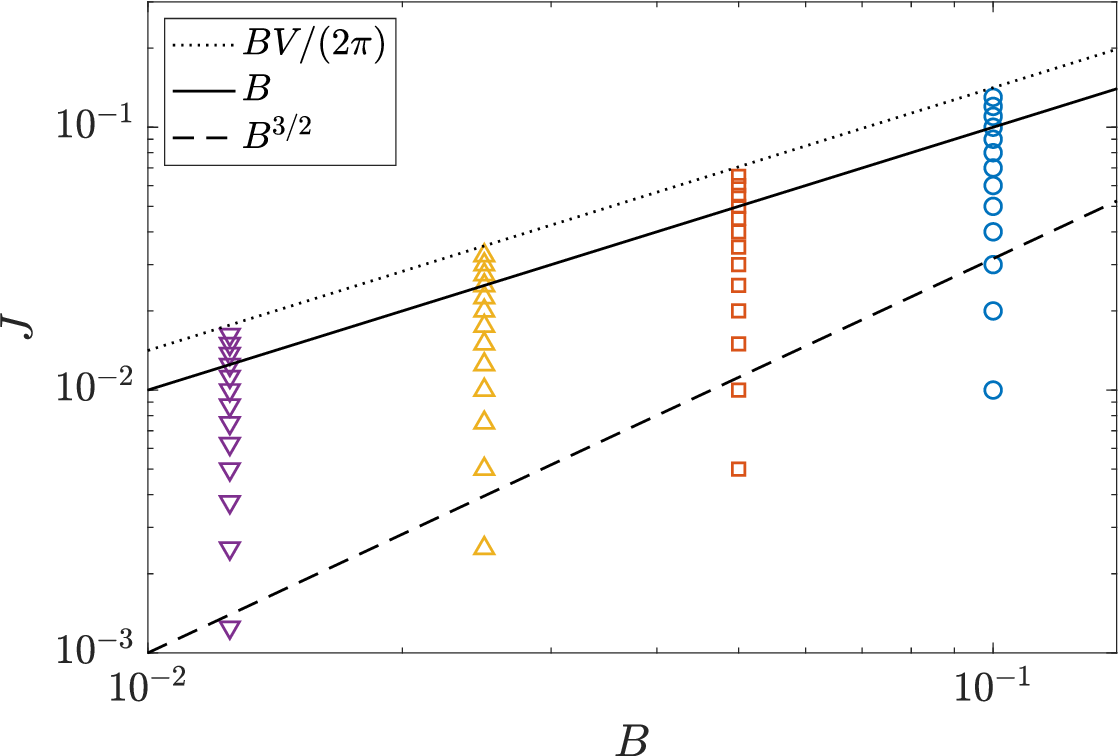}
    \caption{Locations of the numerical simulations in figure \ref{fig:UvsJ} in $J$-$B$ parameter space. Black lines indicate how $J$ scales with $B$ in the theory outlined in \S\ref{sec:asymp_model} (solid), and in the theory discussed in appendix \ref{sec:appJB32} (dashed). The dotted line shows the prediction for the maximum value of $J$ for which collar motion is possible, according to \eqref{globalstressbalanceMATCHED}. }
    \label{fig:app_JBmap}
\end{figure}

Figure \ref{fig:UvsJ}\textcolor{black}{(a)} suggests that the collar speed, $\mathcal{U}$, determined via \eqref{appB32_GSBlinapprox}, provides improved agreement with numerical simulations at small values of $\mathcal{J}=B^{1/2}\hat{J}$, compared to the speed prediction from \eqref{globalstressbalanceMATCHED}. Since the values of $B$ and $J$ used in simulations in figure \ref{fig:UvsJ} are not extremely small, it is not necessarily clear whether the assumption that $J=O(B)$ made in \S\ref{sec:asymp_model}, or the assumption that $J=O(B^{3/2})$ made in this appendix, or another scaling for $J$, is most appropriate for each simulation. Figure \ref{fig:app_JBmap} shows the locations of each of those simulations from figure \ref{fig:UvsJ} in $J$-$B$ space, along with the lines $J=B$ and $J=B^{3/2}$. We can expect those simulations lying close to $J=B$ may obey the $J=O(B)$ asymptotic regime, and those lying close to the line $J=B^{3/2}$ may obey the $J=O(B^{3/2})$ asymptotic regime more closely. However, figure \ref{fig:app_JBmap} indicates that at those values of $B$ used in simulations, there is generally not a great deal of separation between $B$ and $B^{3/2}$, so for many of the simulations it is not immediately clear which regime may provide the better predictions. We can expect that, if much smaller values of $B$ were able to be accessed, simulations with parameter values near the $J=B$ line would obey the $J=O(B)$ theory closely, simulations with parameter values near the line $J=B^{3/2}$ would obey the $J=O(B^{3/2})$ theory closely, and those with parameter values lying far below the line $J=B^{3/2}$ would exhibit essentially Newtonian behaviour. 

\section{Inner regions within the front Bretherton region}\label{sec:inbreth}

\setcounter{equation}{0}
\renewcommand\theequation{B.\arabic{equation}}

In the small-$B$ asymptotic model presented in \S\ref{sec:asymp_model}, the leading-order layer height in the front Bretherton region has an undulating shape, with $\mathcal{H}(\xi)=H_\infty^-$ at a series of points, which we shall call $\xi=\{\xi_0,\xi_1,\xi_2,\dots\}$, with $\xi_0<\xi_1<\dots$. Sufficiently close to each of these points, the expansions \eqref{Breth_expansions} are no longer valid, as the final term in the yield criterion \eqref{YsmallBdefn} enters the leading-order balance. To resolve this, we introduce short inner regions, of length $O(B)$, around each point, $\xi=\{\xi_0,\xi_1,\xi_2,\dots\}$, in which $h\approx BH_\infty^-$ and $Y$ is not necessarily close to $h$. We expect that, within these inner regions, there will be only small changes in $h$ compared to the `outer' Bretherton region solution, i.e. the solution of \eqref{bretherton_eqn1}, but that there may be significant variation in $Y$ within these regions, reflecting the fact that the leading-order shear stress changes sign across each inner region. 

Consider a neighbourhood of $\xi=\xi_n$, for some $n$, and define an inner coordinate, $Z$, such that $\xi = \xi_n+B^{1/2}Z$. Motivated by the form of the `outer' solution around $\xi\approx\xi_n$, we propose that in the inner region,
\begin{equation}
    h = B\mathcal{H}_-(\xi) + B^3\tilde{H}(Z), \quad Y = B\tilde{y}(Z),
    \label{inbreth_expansions}
\end{equation}
where $\mathcal{H}_-(\xi)$ is the solution to \eqref{bretherton_eqn1}. In the neighbourhood of $\xi=\xi_n$, i.e. where $Z=O(1)$ and $|\mathcal{H}_-- H_\infty^-|=O(|\xi-\xi_n|)$, \eqref{bretherton_eqn1} implies that $\mathcal{H}_-'''=O(|\xi-\xi_n|)$, so
\begin{multline}
    \mathcal{H}_- = H_\infty^- + \alpha_n (\xi-\xi_n) + \frac{\beta_n}{2}(\xi-\xi_n)^2 + \frac{\mathcal{U}\alpha_n}{8(H_\infty^-)^3}(\xi-\xi_n)^4 + \dots \\= H_\infty^- + \alpha_n B^{1/2}Z + \frac{\beta_n}{2}BZ^2 + \frac{\mathcal{U}\alpha_n}{8(H_\infty^-)^3}B^2Z^4 + \dots.
    \label{inbreth_outerH}
\end{multline}
Inserting \eqref{inbreth_expansions} and \eqref{inbreth_outerH} into \eqref{travelling_evoleqn} gives
\begin{equation}
    \frac{\mathcal{U}\alpha_n Z}{H_\infty^--\tilde{y}/3} = \frac{1}{2}\tilde{y}^2\left(\frac{3\mathcal{U}
    \alpha_n Z}{(H_\infty^-)^3}+\tilde{H}'''\right),
    \label{inbreth_evoleqn}
\end{equation}
and \eqref{YsmallBdefn} gives
\begin{equation}
    \mathcal{J} = (H_\infty^--\tilde{y})\left|\frac{3\mathcal{U}\alpha_n Z}{(H_\infty^-)^3}+\tilde{H}'''\right|.
    \label{inbreth_Ydefn}
\end{equation}
Combining \eqref{inbreth_evoleqn} and \eqref{inbreth_Ydefn} gives an implicit equation for $\tilde{y}$ within in the inner region,
\begin{equation}
    \mathcal{J}\tilde{y}^2\left(H_\infty^--\frac{\tilde{y}}{3}\right) = |2\mathcal{U}\alpha_n Z|(H_\infty^--\tilde{y}),
    \label{inbreth_Y}
\end{equation}
which can be solved given $\alpha_n = \mathcal{H}_-'(\xi_n)$. From \eqref{inbreth_Y}, 
\begin{equation}
    \tilde{y}\sim H_\infty^- - \frac{\mathcal{J}(H_\infty^-)^3}{3|\mathcal{U}\alpha_n Z|}\quad\mbox{as}\quad |Z|\rightarrow\infty,
\end{equation}
confirming that the inner region solution matches to the leading-order outer solution, in which $Y \sim BH_\infty^-$. Taking the inner limit of \eqref{inbreth_Y} gives
\begin{equation}
    \tilde{y}^2\sim \frac{2|\mathcal{U}\alpha_n Z|}{\mathcal{J}}\quad
    \mbox{as}\quad Z\rightarrow0.
\end{equation}
The yield surface has large slope near $Z=0$, violating the original long-wave assumption, but we do not pursue this potential inconsistency here. 

We expect that there is not necessarily an infinite sequence of inner regions for which this solution holds. Rather, we expect there to be a finite distance at which the Bretherton region solution, outlined in \S\ref{sec:breth}, and hence the inner-region solutions outlined in this section, break down. We expect this to occur when $|\xi|=O(\log({1/B}))$ with $\xi<0$, at which point $|H_\infty^--\mathcal{H}_-|=O(B^{1/2})$ (this can be seen from \eqref{bretherton_eqn1} and the exponential decay of $\mathcal{H}_-$ towards $H_\infty^-$ in the far-field) and then the assumption in \eqref{Breth_expansions} that $Y\approx h$ no longer holds. Therefore, we expect there to be a finite sequence of the inner regions we have outlined here, around points $\xi=\{\xi_0,\xi_1,\dots,\xi_N\}$, for some $N$, before the Bretherton solution breaks down and the fluid transitions towards becoming unyielded. In appendix \ref{sec:rigidif}, we outline how the layer transitions from being in an almost fully-yielded state in the Bretherton regions (both at the front and rear of the collar) to becoming unyielded at a finite distance from the collar.

\section{Rigidification regions}\label{sec:rigidif}

\setcounter{equation}{0}
\renewcommand\theequation{C.\arabic{equation}}

The solutions, $H_\pm$, to \eqref{bretherton_eqn_rescaled} decay exponentially towards $1$ as $X\rightarrow \pm\infty$ in the far fields away from the collar. We expect that the Bretherton region solutions, outlined in \S\ref{sec:breth}, will break down when when $|\mathcal{H}_\pm-H_\infty^\pm| = O(B^{1/2})$, as we would no longer expect that $Y\approx B\mathcal{H}_\pm$ to leading order, as is assumed in \eqref{Breth_expansions}. We have identified this breakdown of the Bretherton region solution in multiple short inner regions in \S\ref{sec:inbreth}, but we expect that it also occurs in other regions close to the points at which the film thickness becomes uniform and $Y\rightarrow0^+$, at the front and rear ends of the collar. We call these regions `rigidification regions', since within them, $Y$ transitions from being close to $h$ to being close to zero. We outline the structures of these regions, at the front and rear of the collar, below. 

At the rear, we suppose that there is a point, $\zeta = \zeta_0$, such that $Y\rightarrow0^+$ as $\zeta\rightarrow \zeta_0^-$ (figure \ref{fig:sketch_structure}c). We then define a local coordinate, $\tilde{z}$, such that $\zeta = \zeta_0 - B^{1/2}\tilde{z}$. If we make the expansions \eqref{Breth_expansions}, we recover the classical equation \eqref{bretherton_eqn1} from the Bretherton region. When $h = B(H_\infty^+ + \tilde{H})$, where $\tilde{H}\ll H_\infty^+$, \eqref{bretherton_eqn1} then suggests exponential dependence of $\tilde{H}$ on $\tilde{z}$. However, from \eqref{YsmallBdefn}, it can be seen that the assumption that $Y\approx h$ is violated when $\tilde{H}=O(B^{1/2})$, which, given the exponential dependence of $\tilde{H}$ on $\tilde{z}$, occurs when $\tilde{z}=O(\log({1/B}))$. Therefore, strictly, we expect the solution outlined in \S\ref{sec:breth} to hold when $|\tilde{z}|\gg\log({1/B})$, and we require a separate inner solution when $|\tilde{z}|\ll\log({1/B})$. This motivates the $O(B^{1/2}\log(1/B))$ length scale identified in the sketch in figure \ref{fig:sketch_structure}(c) for the distance between $\zeta=\zeta_0$ and the Bretherton region. We call the inner region where $\tilde{z}\ll\log(1/B)$ a `rigidification region'.

In the rigidification region at the rear, we assume $|\tilde{z}|\ll\log({1/B})$ and let
\begin{equation}
    h = BH_\infty^+ + B^{3/2}\tilde{H}_+(\tilde{z}) + \dots, \quad Y = B\tilde{Y}_+(\tilde{z}).
    \label{rigidif_expansions}
\end{equation}
Substituting \eqref{rigidif_expansions} into \eqref{travelling_evoleqn} and \eqref{YsmallBdefn} gives
\begin{equation}
    -\frac{1}{2}\tilde{Y}_+^2\tilde{H}_{+,\tilde{z}\tilde{z}\tilde{z}} = \frac{\mathcal{U}\tilde{H}_+}{H_\infty^+ - \frac{\tilde{Y}_+}{3}},
    \label{rigidif_evoleqn}
\end{equation}
and
\begin{equation}
    |\tilde{H}_{+,\tilde{z}\tilde{z}\tilde{z}}|(H_\infty^+ - \tilde{Y}) = \mathcal{J},
    \label{rigidif_Yeqn}
\end{equation}
which could be solved to find $\tilde{H}$ and $\tilde{Y}$ in $\tilde{z} > 0$. However, here, we focus on the inner and outer limits of \eqref{rigidif_evoleqn} and \eqref{rigidif_Yeqn}. In the outer limit, where $\tilde{z}\rightarrow\infty$ and $\tilde{Y}\rightarrow H_\infty^+$, \eqref{rigidif_evoleqn} recovers the behaviour of the Bretherton region solution, with $\tilde{H}$ exhibiting an exponential dependence on $\tilde{z}$, so this inner solution can match adequately onto the main Bretherton region solution. In the inner limit, where $\tilde{z}\rightarrow0$, we expand, and use \eqref{rigidif_evoleqn} and \eqref{rigidif_Yeqn}, to find
\begin{equation}
    \tilde{H}_+ = \frac{\mathcal{J}\tilde{z}^3}{6H_\infty^+} + \dots,\quad \tilde{Y}_+ = \sqrt{\frac{|\mathcal{U}|}{3H_\infty^+}}\,\,\tilde{z}^{3/2} + \dots \quad\mbox{as}\quad \tilde{z}\rightarrow0^+.
    \label{rigidif_innerexpansions}
\end{equation}
This inner limit describes how the fluid matches onto the uniform film at $\zeta=\zeta_0$, with $h_{\zeta\zeta\zeta}$ exhibiting a finite jump dicontinuity as $Y\rightarrow0^+$. We will show in \S\ref{sec:results_smallB} (see figure \ref{fig:structure_comparison}(e) below) that this behaviour is observed in numerical solutions of the thin-film evolution equation \eqref{TFevoleqn}. 

At the front edge, we expect that there is another similar rigidification region between the Bretherton region and the uniform film. We expect that there is a point, $\zeta=\zeta_1$, with $Y\rightarrow0^+$ as $\zeta\rightarrow \zeta_1^+$. Defining the local coordinate now via $\zeta=\zeta_1+B^{1/2}\tilde{z}$, we again identify that the Bretherton region solution from \S\ref{sec:breth} holds when $|\tilde{z}|\gg\log(1/B)$. In the rigidification region, we assume $|\tilde{z}|\ll\log(1/B)$, and let
\begin{equation}
    h = BH_\infty^- + B^{3/2}\tilde{H}_-(\tilde{z}) + \dots, \quad Y = B\tilde{Y}_-(\tilde{z}),
    \label{rigidif_expansions2}
\end{equation}
which gives
\begin{equation}
    \frac{1}{2}\tilde{Y}_-^2\tilde{H}_{-,\tilde{z}\tilde{z}\tilde{z}} = \frac{\mathcal{U}\tilde{H}_-}{H_\infty^- - \frac{\tilde{Y}_-}{3}},
    \label{rigidif_evoleqn2}
\end{equation}
and
\begin{equation}
    |\tilde{H}_{-,\tilde{z}\tilde{z}\tilde{z}}|(H_\infty^- - \tilde{Y}) = \mathcal{J},
    \label{rigidif_Yeqn2}
\end{equation}
Again, rather than solving the equations \eqref{rigidif_evoleqn2} and \eqref{rigidif_Yeqn2} in full, we focus on their inner and outer limits. In the outer limit of the rigidification region, where $|\tilde{z}|\rightarrow\infty$, we again recover the exponential dependence of $\tilde{H}$ on $\tilde{z}$, so that this solution can, in theory, be matched onto the undulating solution for $h$ in the Bretherton region. However, at the front, we do not expect the inner limit of the rigidification region to have the structure \eqref{rigidif_innerexpansions}. Instead, we expect a more intricate structure in the rigidification region, with $\tilde{H}$ potentially undulating and repeatedly changing signs. In the inner limit, where $\tilde{z}\rightarrow0^+$ and $\tilde{Y}\rightarrow0^+$, \eqref{rigidif_Yeqn2} reduces to
\begin{equation}
    \tilde{H}_{-,\tilde{z}\tilde{z}\tilde{z}} = -\frac{\mathcal{J}\mathrm{sgn}(\tilde{H}_-)}{H_\infty^-},
    \label{rigidif_front_ineqn}
\end{equation}
where we have also used \eqref{rigidif_evoleqn2}, and the fact that $\mathcal{U}\leq0$, to reach \eqref{rigidif_front_ineqn}. The structure of the solution to \eqref{rigidif_front_ineqn} was detailed by Jalaal et al. \cite{jalaal_stoeber_balmforth_2021}, since the same equation arises at the leading edge of a droplet spreading into a uniform precursor film. \textcolor{black}{The same equation also appears when considering long bubbles translating in tubes, providing the solution very close to where the fluid rigidifies in the uniform film deposited on the tube wall \cite{jalaal_long_2016}. The difference in the present analysis is that, since we focus on the small-$B$ limit with $\mathcal{J}=O(1)$, the rigidification region is always adjacent to an almost fully-yielded `Bretherton' region. }

\textcolor{black}{As explained by Jalaal et al. \cite{jalaal_stoeber_balmforth_2021}, }the solution for $\tilde{H}_-$ in \eqref{rigidif_front_ineqn} is composed of an infinite sequence of cubic polynomials, with the $n^{\mathrm{th}}$ cubic being defined in an interval, $I_n=[\tilde{z}_n,\tilde{z}_{n+1}]$, and continuity of $\tilde{H}_-$ and its first and second derivatives being enforced at the boundaries between each of the intervals $\{I_n\}$. The third derivative is not, in general, continuous at $\tilde{z}=\tilde{z}_n$. This results in an undulating structure for $\tilde{H}_-$, with both the amplitude and wavelength of the undulations decreasing as $n\rightarrow\infty$, with the train of undulations ending at a finite distance \cite{jalaal_stoeber_balmforth_2021}, which we can identify as the point $\tilde{z}=0$. Again, this structure is inconsistent (over very small scales) with the long-wave assumption underpinning the original model, but we do not pursue this detail here.  

The wall shear stress in each of the rigidification regions is
\begin{equation}
    |\tau_\mathrm{w}| \sim BH_\infty^\pm \tilde{H}_{\pm,\tilde{z}\tilde{z}\tilde{z}},
\end{equation}
which when integrated over the length of the regions gives an $O(B^{3/2})$ contribution to the total viscous drag exerted on the collar. This is not a leading-order contribution, so the structure of the rigidification regions does not influence the leading-order global force balance \eqref{globstressbal} or the calculation of the translation speed, $\mathcal{U}$. However, the emergence of these rigidification regions provides further insight into the structure of the asymptotic solution: in the Newtonian problem, there is an infinitely long train of undulations in the front Bretherton region, and an exponential approach to the uniform film in the rear Bretherton region, but in the viscoplastic problem, the Bretherton regions are truncated at a distance $O(B^{1/2}\log(1/B))$ from the collar body, at which point the liquid becomes unyielded and the film thickness becomes uniform. 

\section{Higher-order corrections to the small-$B$ asymptotic solution}\label{sec:appHOcorrections}
\setcounter{equation}{0}
\renewcommand\theequation{D.\arabic{equation}}

In this appendix, we seek higher-order corrections to the asymptotic model presented in \S\ref{sec:asymp_model} in order to determine a correction to the collar speed, $\mathcal{U}$. We extend the analysis presented in \S\ref{sec:asymp_main_body} and \S\ref{sec:asymp_int} to determine a higher-order correction to the wall shear stress, $\tau_{\mathrm{w}}$, across the collar. The global force balance \eqref{globstressbal} can then be combined with this adjusted wall shear stress to arrive at a modified approximation for $\mathcal{U}$. 

In the main body of the collar, we propose expansions,
\begin{equation}
    h = h_0(\zeta) + B\log{B}h_1(\zeta) + Bh_2(\zeta) + B^{5/4}h_3(\zeta) + \dots, \quad Y = B^{1/4}Y_0(\zeta) + B^{1/2}Y_1(\zeta) + \dots,
    \label{HOTbodyexpansions}
\end{equation}
where $h_0$, $h_1$, $h_2$ and $Y_0$ are given in \eqref{body_h0}, \eqref{body_h1}, \eqref{body_h2} and \eqref{body_Y0}, respectively. Inserting \eqref{HOTbodyexpansions} into \eqref{travelling_evoleqn} gives
\begin{equation}
    0 = \frac{1}{2}h_0Y_0^2(h_3'+h_3''') + (1-h_2'-h_2''')\left[-h_0Y_0Y_1 + \frac{1}{6}Y_0^3\right],
    \label{app_evoleqn}
\end{equation}
and \eqref{YsmallBdefn} gives
\begin{equation}
    h_3'+h_3''' = -\frac{\mathcal{J}Y_0}{h_0^2}. 
    \label{app_Yeqn}
\end{equation}
Combining \eqref{app_evoleqn} and \eqref{app_Yeqn}, and using \eqref{body_Yeqn}, gives
\begin{equation}
    Y_1 = -\frac{2|\mathcal{U}|}{3\mathcal{J}}.
    \label{app_Y1}
\end{equation}
The inclusion of the first-order correction \eqref{app_Y1} can improve the agreement between the asymptotic prediction for $Y$ and the computed solution in numerics (figure \ref{fig:structure_comparison}a). 

Expanding the wall shear stress \eqref{asymp_tauw}, in the main collar body, and using \eqref{app_Yeqn}, gives
\begin{equation}
    \tau_\mathrm{w} = -B\mathcal{J} - B^{5/4}\sqrt{\frac{2|\mathcal{U}|\mathcal{J}}{h_0}} + \dots.
    \label{app_tauw}
\end{equation}
By integrating $\tau_{\mathrm{w}}$ across the length of the collar, we can approximate the correction to the viscous drag exerted at the wall. From \eqref{app_tauw}, 
\begin{equation}
    \int_{\zeta_c}^{2\pi-\zeta_d}(-\tau_\mathrm{w})\,\mathrm{d}\zeta = 2\pi \mathcal{J}B + B^{5/4}\int_{\zeta_c}^{2\pi-\zeta_d}\sqrt{\frac{4\pi|\mathcal{U}|\mathcal{J}}{V(1-\cos\zeta)}}\,\mathrm{d}\zeta + \dots,
    \label{stressint1}
\end{equation}
where we take $\zeta_c = B^\alpha\bar{\zeta}_c$ and $\zeta_d = B^\alpha\bar{\zeta}_d$, where $0<\alpha<\tfrac{1}{4}$, so that in \eqref{stressint1} we have integrated up to the intermediate regions. Then, evaluating \eqref{stressint1}, we find
\begin{equation}
    \int_{\zeta_c}^{2\pi-\zeta_d}\sqrt{\frac{4\pi|\mathcal{U}|\mathcal{J}}{V(1-\cos\zeta)}}\,\mathrm{d}\zeta \sim 4\alpha\sqrt{\frac{2\pi|\mathcal{U}|\mathcal{J}}{V}}\log\left(1/B\right) + O(1).
    \label{stressint2}
\end{equation}
Combining \eqref{stressint1} and \eqref{stressint2}, 
\begin{equation}
    \int_{\zeta_c}^{2\pi-\zeta_d}(-\tau_\mathrm{w})\,\mathrm{d}\zeta = B\mathcal{J}(2\pi -B^{\alpha}\bar{\zeta}_c-B^{\alpha}\bar{\zeta}_d) + B^{5/4}\log\left(1/B\right)4\alpha\sqrt{\frac{2\pi|\mathcal{U}|\mathcal{J}}{V}} + O(B^{5/4}).
    \label{stressint3}
\end{equation}
The factor $\alpha$ is undetermined, and to fully determine the first correction to the wall shear stress integral across the collar, we must consider the contribution from the intermediate regions. 

In the intermediate regions, as in \S\ref{sec:asymp_int}, we propose the expansions,
\begin{eqnarray}
    h &=& B^{1/2}\frac{V}{4\pi}\hat{X}^2 + B\left(\hat{A}_\pm - \frac{V\hat{X}^4}{48\pi}\right) + B^{5/4}\hat{h}_{\pm} + \dots,\label{app_int_hexpansion}\\
    Y &=& B^{1/2}\hat{Y}_{\pm} + \dots,\label{app_int_Yexpansion}
\end{eqnarray}
where $\hat{h}_{\pm}(\hat{X})$ and $\hat{Y}_{\pm}(\hat{X})$ are determined via \eqref{intermediate_evoleqn} and \eqref{intermediate_Yeqn}. In the limits of the intermediate regions approaching the main collar body, \eqref{intermediate_evoleqn} and \eqref{intermediate_Yeqn} give
\begin{eqnarray}
    \hat{Y}_\pm\sim \sqrt{\frac{|\mathcal{U}|V}{2\pi\mathcal{J}}}|\hat{X}| -\frac{2|\mathcal{U}|}{3\mathcal{J}} + \dots \quad\mbox{as}\quad \hat{X}\rightarrow\mp\infty,\\
    \hat{h}'''_\pm \sim -\frac{4\pi\mathcal{J}}{V\hat{X}^2} - \frac{8}{|\hat{X}|^3}\sqrt{\frac{2\pi^3|\mathcal{U}|\mathcal{J}}{V^3}}+\dots
    \quad\mbox{as}\quad \hat{X}\rightarrow\mp\infty,
    \label{appA_hhatlims}
\end{eqnarray}

In the intermediate region, the wall shear stress at leading order is
\begin{equation}
    \tau_{\mathrm{w}}\sim B\frac{V\hat{X}^2}{4\pi}\hat{h}_{\pm}'''+\dots.\label{app_int_tau0}
\end{equation}
From \eqref{appA_hhatlims} and \eqref{app_int_tau0}, we expect the integral of $\tau_\mathrm{w}$ to be logarithmic as the intermediate region solution approaches the main collar body, so we expect that the dominant contribution to the viscous drag in the intermediate regions arises from the outer limits of the intermediate regions where they approach the main collar body. Therefore, to find the leading-order contribution to the integral of the wall shear stress from the intermediate regions, we consider the limit of \eqref{app_int_tau0} as the intermediate region approaches the main collar body. From \eqref{appA_hhatlims},
\begin{equation}
    -\frac{V\hat{X}^2}{4\pi}\hat{h}_{\pm}'''\sim \mathcal{J} + \frac{1}{|\hat{X}|}\sqrt{\frac{8\pi|\mathcal{U}|\mathcal{J}}{V}} + \dots \quad\mbox{as}\quad \hat{X}\rightarrow\mp\infty,\label{app_int_tauexpansion}
\end{equation}
and if we evaluate the integral of $\tau_{\mathrm{w}}$ by using the expression \eqref{app_int_tauexpansion}, we can recover the leading-order contribution to the wall shear stress integral. The leading-order integral across the front intermediate region, using \eqref{app_int_tauexpansion}, is 
\begin{multline}
    \int_{\mathrm{int}_-}(-\tau_\mathrm{w})\,\mathrm{d}\zeta \sim 
    \int_{B^{1/4}X_0}^{B^\alpha\bar{\zeta_c}}\left(-\tau_{\mathrm{w}}\right)\,\mathrm{d}\zeta \sim
    B^{5/4}\int_{X_0}^{B^{(\alpha-1/4)}\bar{\zeta_c}}\left(\mathcal{J} + \frac{1}{|\hat{X}|}\sqrt{\frac{8\pi|\mathcal{U}|\mathcal{J}}{V}}\right)\,\mathrm{d}\hat{X}
    \\
    \sim B^{1+\alpha}\mathcal{J}\bar{\zeta}_c + B^{5/4}\log{B}\left(\alpha-\frac{1}{4}\right)\sqrt{\frac{8\pi|\mathcal{U}|\mathcal{J}}{V}} + O(B^{5/4}),\label{app_int_fronttauint}
\end{multline}
where $X_0=O(1)$ is some constant, so $\zeta=B^{1/4}X_0$ is some point in the interior of the intermediate region. By symmetry, the integral across the rear intermediate region will be the same as \eqref{app_int_fronttauint} but with $\bar{\zeta}_c$ replaced by $\bar{\zeta}_d$. Therefore, summing the integrals of $\tau_\mathrm{w}$ across the two intermediate regions and the main collar body \eqref{stressint3} gives
\begin{equation}
    \int_{B^{1/4}X_0}^{2\pi-B^{1/4}X_0}(-\tau_\mathrm{w})\,\mathrm{d}\zeta \sim BD_0 + B^{5/4}\log{B}D_1
     + O(B^{5/4}).
    \label{app_tauinttotal}
\end{equation}
where
\begin{equation}
    D_0 = 2\pi\mathcal{J}\quad\mbox{and}\quad D_1 = -\sqrt{\frac{2\pi|\mathcal{U}|\mathcal{J}}{V}}.
\end{equation}

The expression \eqref{app_tauinttotal} contains both the leading-order viscous drag across the collar and the first-order correction. Replacing the left-hand-side of \eqref{globalstressbalanceMATCHED}, i.e. the leading-order viscous drag on the collar, with \eqref{app_tauinttotal} gives a composite approximation for the global force balance,
\begin{equation}
    BD_0 + B^{5/4}\log{B}D_1
    \approx
    BV + B(3|\mathcal{U}|)^{2/3}(G_0G_2-H_0H_2).
    \label{app_GSBcomp}
\end{equation}
We can approximate $D_1$ by using the expression for collar speed determined from leading-order theory \eqref{globalstressbalanceMATCHED}, then determine the adjusted expression for $\mathcal{U}$ via \eqref{app_GSBcomp}; this results in generally improved agreement between asymptotic predictions for collar speeds and collar speeds in numerical simulations (figure \ref{fig:UvsJ}\textcolor{black}{a}).

\bibliography{bibliography}

\end{document}